\documentclass[fleqn,usenatbib]{mnras}

\usepackage{newtxtext,newtxmath}

\usepackage[T1]{fontenc}
\usepackage{enumitem}
\usepackage[normalem]{ulem}
\DeclareRobustCommand{\VAN}[3]{#2}
\let\VANthebibliography\thebibliography
\def\thebibliography{\DeclareRobustCommand{\VAN}[3]{##3}\VANthebibliography}

\usepackage{graphicx}	
\usepackage{amsmath}	

\title[Disk Survival Time-scales]{Protoplanetary Disk Survival Time-scales: A Blind Survey of Young Clusters up to 100 Myr in the Solar Vicinity}

\author[G. M. Ben et al.]{
Gregory Mathews Ben,$^{1}$\thanks{E-mail: gregorygmb10@gmail.com}
Jessy Jose,$^{1}$\thanks{E-mail: jessyvjose1@gmail.com}
Jes\'us Hern\'andez $^{2}$
\\
$^{1}$Department of Physics, Indian Institute of Science Education and Research Tirupati, Yerpedu, Tirupati - 517619, Andhra Pradesh, India\\
$^{2}$Instituto de Astronom\'ia, Universidad Nacional Aut\'onoma de M\'exico, Apartado Postal 106, C. P. 22800, Ensenada, B. C., M\'exico\\
}

\date{Accepted XXX. Received YYY; in original form ZZZ}

\pubyear{\the\year{}}

\begin{document}
\label{firstpage}
\pagerange{\pageref{firstpage}--\pageref{lastpage}}
\maketitle

\begin{abstract}
We study the protoplanetary disk lifetimes using a large sample of young stellar objects in nearby clusters. To investigate the final phase of disk dissipation, we selected 32 clusters,  located within 500 pc and aged between 1 and 100 Myr, with membership determined using Gaia data. The Age and mass information of the sources are obtained through spectral energy distribution (SED) analysis 
and using evolutionary models of various ages. Using the IR data from 2MASS and WISE catalogues, we employ three methods to identify disks across the different wavelength regimes (1.1- 22 $\mu$m). We find that disk fraction consistently decreases as stellar systems age, a trend observed across all wavelengths included in this study. However, there is an increase in the time scale of disk decay as wavelength increases, with characteristic timescales of $\tau_{\text{short}}$ = 1.6 ± 0.1 Myr for shorter wavelengths (1.6-4.6 $\mu$m) versus $\tau_{\text{W3}}$ = 4.4 ± 0.3 Myr for 12 $\mu$m. This supports the idea that outer disk regions evolve more slowly. Notably, we detect infrared excesses at 12 $\mu$m and 22 $\mu$m in relatively older systems ($>$10 Myr), with some disks with estimated ages up to $\sim$ 100 Myr.
 Among these, we identify a population of full disks that persist beyond the typical dissipation timescale. We also observe that the median mass of disk-hosting stars decreases from 0.62 $M_\odot$ to 0.27 $M_\odot$ in clusters younger and older than 40 Myr, respectively, indicating slower disk dissipation around lower-mass stars. We identify 33 transitional disk candidates using various color-color diagrams. Using LAMOST DR8 optical spectra and H-alpha equivalent widths, we identify possible accretors and estimate their mass accretion rates, finding most are younger than 10 Myr.
\end{abstract}




\section{Introduction}

Understanding the evolution of stars and planets requires a focus on protoplanetary disks orbiting young stellar objects (YSOs). During their lifespan, the gas and dust-filled disks change significantly owing to different processes. Several important mechanisms drive the disk's dissipation as the system evolves: (i) accretion onto the central protostar, which is channeled by magnetic field lines from the inner edge of the disk to the stellar surface \citep{2007prpl.conf..479B,2016ARA&A..54..135H}; (ii) accumulation of material into formation of planets (\citealt{2023ASPC..534..501M}, and references therein); (iii) photoevaporation caused  by the protostar's radiation \citep{2014prpl.conf..451F,2019MNRAS.487..691P,2023ASPC..534..567P}; (iv) external photoevaporation due to nearby massive stars \citep{2006ApJ...641..504A,2016MNRAS.457.3593F,2018MNRAS.478.2700W,2024MNRAS.535.1321D,2025A&A...693A..87M}; (v) dynamical disk dissipation produced by giant planets or stellar binary systems \citep{2011ApJ...732...42A,2012ApJ...745...19K} ; and (vi) 
stellar flybys can alter disk morphology and cause significant mass loss, particularly in dense clusters where stellar encounters are expected
\citep{1996MNRAS.278..303H,2011MNRAS.411..859M,2014A&A...564A..28P}.

The identification of protoplanetary disks is usually achieved by detecting excess infrared (IR) emissions that exceed the amount released by the star's photosphere. The star's light heats the dust in the disk, causing excess infrared radiation that is re-emitted in the near to far-infrared and sub-mm wavelengths. Different wavelengths probe different sizes of dust particles and disk regions around stars. 
Sub-$\mu$m particles that efficiently contribute to scatter lights grow about 14 orders of magnitude for building planets \citep[e.g.,][]{Armitage2013}. The near-IR wavelengths trace the inner part of the disk, mainly the inner dust sublimation wall produced by stellar radiation. Longer wavelengths (e.g., mid- and far- IR) trace the thermal emission of regions located at tens of astronomical units (AU) from the star. The sub-mm radiation traces the outer disk and the mid plane of primordial disks, and for debris disks, this radiation traces larger grains marking the positions of parent bodies whose collisions yield the dust seen at shorter wavelengths \citep{2017ApJ...836...34M}.
Additionally, young disks are actively transferring mass onto the stellar surface, with the resulting accretion shocks releasing energy. These accretion shocks appear as a continuum excess dominating the observed emission in both the ultraviolet continuum and emission line features \citep{1998ApJ...509..802C,2016ARA&A..54..135H}.

Our understanding of protoplanetary disks has been greatly advanced by observations across multiple wavelength regimes using various telescopes. High-resolution observations from telescopes such as optical imaging with HST \citep{2014AJ....148...59S,2021MNRAS.504.3074W},  near-IR imaging with JWST, Gemini \citep{Lawson_2023,2021MNRAS.504.3074W} have facilitated
the analysis of  the size and composition of the circumstellar dust in scattered light. ALMA's sub-mm observations have examined the disk demographics and planet formation \citep{2020A&A...642A.164V,2023A&A...673A..77R}. Several infrared telescopes, including {\it IRAS}, {\it WISE}, {\it Spitzer}, {\it Herschel}, {\it UKIDSS} and others, have made substantial contributions to improving our understanding of disk development through their individual observations and surveys \citep{2004ARA&A..42..685Z,2014prpl.conf..195D,2014ApJ...791..131K,Dunham_2015,2024MNRAS.535.1321D}.

Several observational analyses show that, within 2-6 Myr
less than about 10\%  of stars retain their disks, and almost
all stars lose their disks by approximately 10 Myr \citep{2001ApJ...553L.153H,Richert_2018}. However, the majority of these studies have focused mainly on young clusters of < 10 Myr \citep{2001ApJ...553L.153H,2006ApJ...638..897S,2007ApJ...662.1067H,2021A&A...650A.157G,2021MNRAS.500.3123D,2024ApJ...970...88P}, and relatively older clusters (i.e., > 10 Myr) being generally under-represented.  Recent works by Pfalzner \citep{2022ApJ...939L..10P,2024ApJ...963..122P} demonstrate that the disk lifetimes for M-type stars are more widely distributed, with median lifetimes of 5–10 Myr and some disks persisting beyond 20 Myr. Importantly, \cite{2022ApJ...939L..10P} suggest that there is an underestimate of low-mass disk stars due to completeness issues in distant regions, implying that past surveys may have overlooked a significant population of long-lived disks around low-mass stars.  This is further highlighted by the discovery of "Peter Pan" disks, which are protoplanetary disks that continue to accrete for considerably longer periods than previously thought feasible, for more than 10 Myr \citep[$\sim$45 Myr;][]{2018MNRAS.476.3290M,Silverberg_2020}. These long-lived disks emphasize the need to study disk persistence in low-mass stars in more evolved systems, since no definite age limit indicates the end of the dissipation process. The existence of Peter Pan disks gives clear evidence for lengthy periods of disk accretion, significantly altering our understanding of disk longevity and its implications for planet formation. Thus, there is a clear need to survey a wider range of stellar ages and masses to fully understand disk evolution.

To address this, we conducted an unbiased survey of young stellar clusters across a broader age range with a particular focus on low-mass stars in evolved clusters and located within the solar neighborhood (distance < 500 pc). Previous surveys utilizing IR telescopes (e.g., Spitzer) were 
constrained by their small field of view (FOV), leading to membership constraints, especially for large regions in the sky \citep{2022A&A...664A..66M}. Particularly for centrally concentrated stellar clusters, these small FOVs likely underestimate disk fractions by only observing the central, densely populated regions where disks dissipate faster due to strong stellar interactions and or massive stellar feedback, while missing the expanded outer regions where disks may persist longer  \citep{2014ApJ...793L..34P,2023JApA...44...77D,2024MNRAS.535.1321D, 2024MNRAS.528.5633G}. Our approach overcomes these limitations by employing the all sky surveys including Gaia mission \citep{2016A&A...595A...1G},  2MASS \citep{2006AJ....131.1163S} and WISE \citep{2010AJ....140.1868W}, which offers the possibility to perform membership and disk analysis without any spatial constraints. We use spectroscopic data from the LAMOST survey for mass accretion rate analysis, which gives wide-field coverage. This multi-wavelength technique enables us to transcend the spatial limitations of previous studies.

The structure of this paper is as follows. Section \ref{sec:data} details the sample clusters we have used in our analysis. Section \ref{sec:param} describes the estimation of stellar parameters of the sample clusters. Section \ref{sec:df} deals with the disk fraction analysis. The H$\alpha$ equivalent width analysis and the determination of mass accretion rates are explained in Section \ref{sec:accr}. Various results are discussed in Section \ref{sec:dis}.
\section{Sample clusters and Data sets used}

This section deals with the details of the sample of young clusters in this study and the various data sets used for the analysis. We use the optical and NIR photometry  and optical spectroscopic data from the archive, wherever available. 

\label{sec:data}
\subsection{Young clusters and their membership}
 \cite{2020A&A...640A...1C} [hereafter CG2020] have compiled a membership list of the clusters in the Milky Way using Gaia DR2 data. They used probability analysis, specifically selecting those  stars with
a probability exceeding 70\% and with a brightness greater than G=18 mag for membership analysis. CG2020 employed a trained Artificial Neural Network (ANN) to estimate the clusters' age, distance modulus, and interstellar extinction based on their members' Gaia photometry and their mean Gaia parallax and proper motion. The training dataset was developed using observed clusters with  reliable parameters. This extensive research resulted in a membership list of 1867 clusters covering an age range from 1 Myr to 8 Gyr and distances up to 12 kpc, providing a robust sample for exploring cluster features across different evolutionary stages and Galactic environments.

For our analysis, we selected the clusters with members greater than 100 and with distances less than 500 pc from the list given in CG2020. We imposed these selection criteria to have a statistically significant number of sources in each cluster and  sufficient stellar mass completeness of the members (see section \ref{sec:massd} for details). We also restricted our selection to clusters with ages less than 100 Myr, as beyond this age, it is unlikely to have a significant number of sources with disks around them. Thus, we have a total of 32 clusters for the analysis. The list of clusters is given in Table \ref{table:main}, along with their coordinates, distance and extinction as per CG2020. We examined the parallax and proper motion values of a few well-known young clusters in the list, such as IC 348 and Collinder 69. Our analysis confirms consistency with the mean values reported in CG2020, validating the membership list. Later, we performed our own age analysis to maintain uniformity across the sample, and the details are explained in Section \ref{sec:param}.
Various data sets used for the analysis are described below. 
\subsection{IR photometry: from 2MASS and WISE}
The corresponding IR photometry for the above Gaia-based members for each cluster was obtained from 2MASS \citep{2006AJ....131.1163S} 
 and ALLWISE
 \citep{2010AJ....140.1868W}   catalogs by giving a search radius of 1" around each member. The distance cut-off of 500 parsecs in the above selection was given 
considering the shallow observations of WISE. Owing to the quality of photometric data, we retained
those sources with errors less than 0.2 mag across all the bands in 2MASS and WISE. All 32 clusters in our final sample from the CG2020 list show detection rates of 80\% or above in each of the following bands: 2MASS J, H, K, and WISE W1, W2. We noticed that the detections were significantly lower in longer wavelengths of W3 and W4 bands and we used the photometry in these bands wherever available. This is further dealt with in Section \ref{sec:w3,w4}. We also examined the quality and contamination flags in the WISE bands, confirming that all sources met the photometric quality criteria (ph\_flag = A or B). 
We visually inspected the WISE images of sources with potential contamination indicated by the contamination and confusion flag (e.g., cc\_flags$\neq$0). We found that the detection in many of these sources is real without any evident contamination, as indicated by the lowercase letter in the cc\_flag.
So, we included them in the disk fraction analysis.  In summary, we have 32 young clusters selected for the analysis, having 6856 sources in total. The list of clusters is given in Table \ref{table:main}, along with their various physical parameters. 
\subsection{Optical Spectroscopy from the archive of  LAMOST}
\label{sec:lamost}
In order to survey the accretion signatures among the candidate young sources selected in our analysis, optical spectroscopic data was used. 
The optical spectra were obtained from the archive of the Large sky Area Multi-Object fiber Spectroscopic
Telescope (LAMOST) \citep{2012arXiv1206.3569Z,2012RAA....12.1197C} DR8 low-resolution spectra (R$\approx$1500). Since we are analyzing the $H\alpha$ equivalent widths as the potential signature of accretion among young stellar objects (YSOs), we retained the spectra of sources with Signal-to-Noise ratio (SNR) in SDSS r-band value greater than 10 \citep{2016RAA....16..138H}. This filtration yields 773 sources with optical spectra. The analysis of H$\alpha$ equivalent widths and subsequent follow-up procedures are detailed in Section \ref{sec:accr}.
\section{Physical Parameters of the sample clusters}
\label{sec:param}

SED (Spectral Energy Distribution) fitting is a valuable technique for estimating stellar parameters such as effective temperature and luminosity of the stars \citep{2006A&A...450..735M,2016ApJ...831L...6S,2020ApJ...892..122J,2023JApA...44...77D,2024ApJ...970...88P}. We performed SED fitting using an online tool, VO SED Analyser (VOSA) \footnote{\url{http://svo2.cab.inta-csic.es/theory/vosa/}} \citep{refId0}. VOSA constructs SEDs by compiling photometric data from various online catalogs, including 2MASS, UKIDSS, WISE, Spitzer, GAIA, IPHAS, and PANSTARRS, which are all accessible through VO services. We supply the coordinates, extinction, distance, and the associated error for each source as input parameters, and VOSA then utilizes the source coordinates to search the various catalogs above and retrieve the necessary photometry data wherever available. VOSA utilizes an automatic algorithm to identify infrared excess among candidates, and it only employs photometric bands up until the point where the excess is detected for the SED fitting. For dereddening, we treated the extinction value within a range of $\pm$ 1 magnitude from the cluster's \(A_V\) (Table \ref{table:main}). After correcting for distance and extinction, VOSA would then select SED with the best-fit \(A_V\) value. Details are given in \cite{2021MNRAS.504.2557D} and \cite{2024ApJ...970...88P}.

The observed SED is then compared with synthetic photometry obtained from theoretical BT-Settl (CIFIST) models  \citep{2013MSAIS..24..128A,2011SoPh..268..255C} for the solar metallicity. The best-fitting model is determined using the chi-square minimization technique, and we obtain the temperature, luminosity, and model flux values as the output for each candidate.

\subsection{Age and Mass}
\label{sec:agemass}
\begin{figure}

\includegraphics[width=\columnwidth]{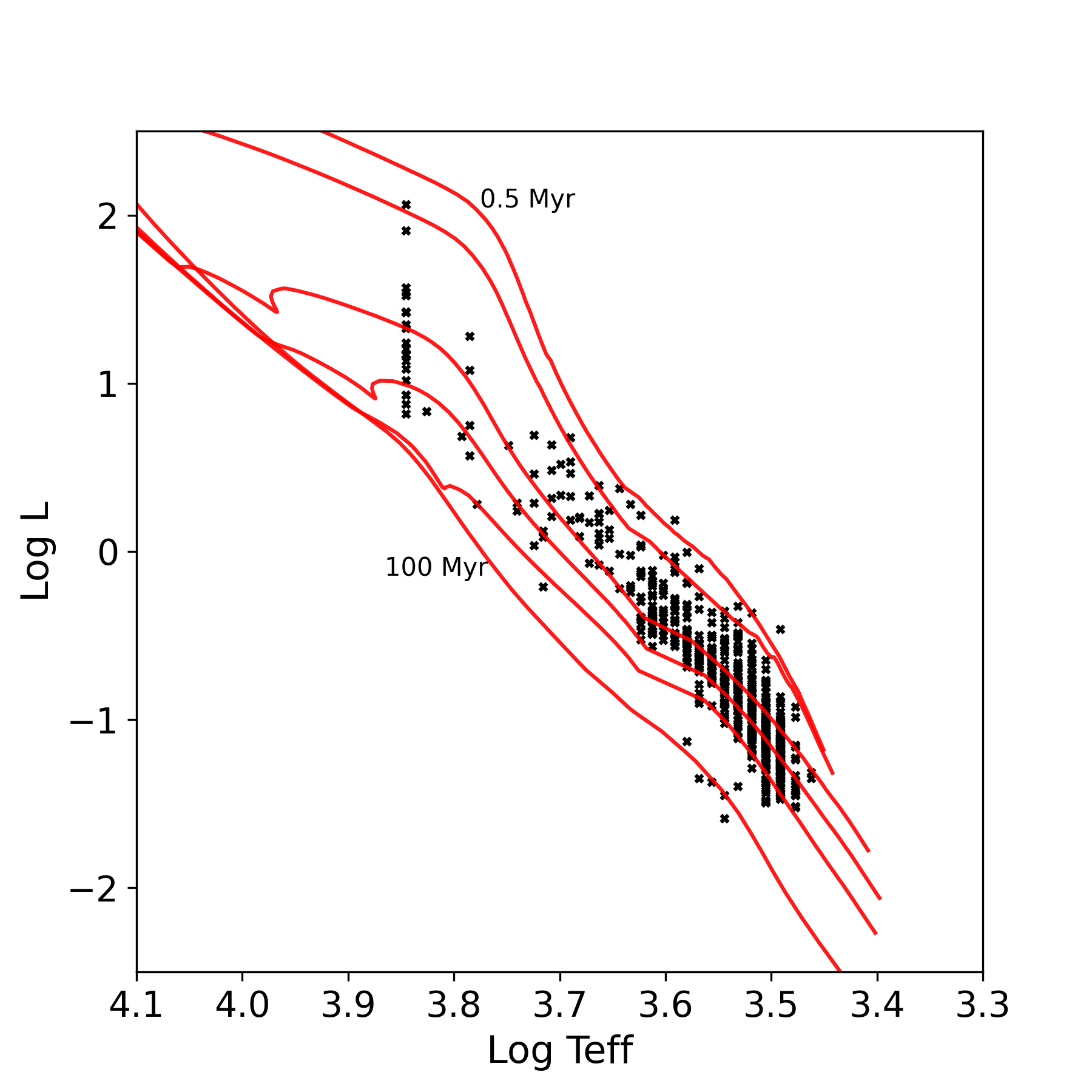}
\caption{HR diagram of candidate members (black crosses) in a sample cluster Collinder 69. The red curves represent isochrones from the PARSEC models, arranged from top to bottom in increasing age: 0.5 Myr, 1 Myr, 5 Myr, 10 Myr, 20 Myr, and 100 Myr. 
}
\label{fig:c69hr}
\end{figure}

Determining the age and mass distribution of young clusters is challenging because these fundamental parameters are highly uncertain and difficult to constrain, especially when dealing with very young clusters \citep{2014prpl.conf..219S,2017ApJ...842..123F,2017ApJ...836...98J}. While it is a common simplification to assume that all stars within a given cluster formed simultaneously from the interstellar medium, numerous studies have revealed that young star-forming regions frequently exhibit non-coeval stellar evolution. This non-coeval nature of star formation leads to a spread in the age estimation as observed in several star-forming regions \citep{2016ApJ...822...49J,2017ApJ...838..150K,2017MNRAS.468.2684P,2021MNRAS.508.3388G,2023ApJ...948....7D,2024MNRAS.528.5633G}. We can estimate  the age and mass by using the Hertzsprung-Russell (HR) diagram. This approach involves plotting cluster members on an HR diagram and comparing them to theoretical pre-main-sequence (PMS) evolutionary tracks and isochrones. The stellar ages and masses can then be estimated by comparing their observed positions to those of theoretical models \citep{Herczeg_2015,2020ApJ...892..122J,2021MNRAS.504.2557D,2023ApJ...948....7D,2025arXiv250316205D}.
We employed the PAdova and TRieste Stellar Evolution Code (PARSEC 1.2) isochrones \citep{2012MNRAS.427..127B} of age between 0.5- 100 Myr  for solar metallicity. The HR diagram for the sources in the sample cluster Collinder 69 is shown in Figure \ref{fig:c69hr}. The PARSEC 1.2 isochrones are shown in red, ranging from 0.5 Myr as the topmost curve to 1, 5, 10, 20, and 100 Myr towards the bottom. The black crosses represent the positions of the sources in the HR diagram.

Each source was assigned the closest matching age and mass by comparing its measured temperature and luminosity from the SED fitting to the intrinsic values provided by the isochrones. We retained only those sources where the difference between the observed and intrinsic temperatures fell within $\pm$ 200 K. 
Since this study focuses mainly on low-mass stars, we included only those with masses less than 2 \(M_\odot\) for the follow-up analysis. The lower mass limit implemented in the analysis is explained in section \ref{sec:massd}.  

\begin{figure}
\includegraphics[width=1\columnwidth]{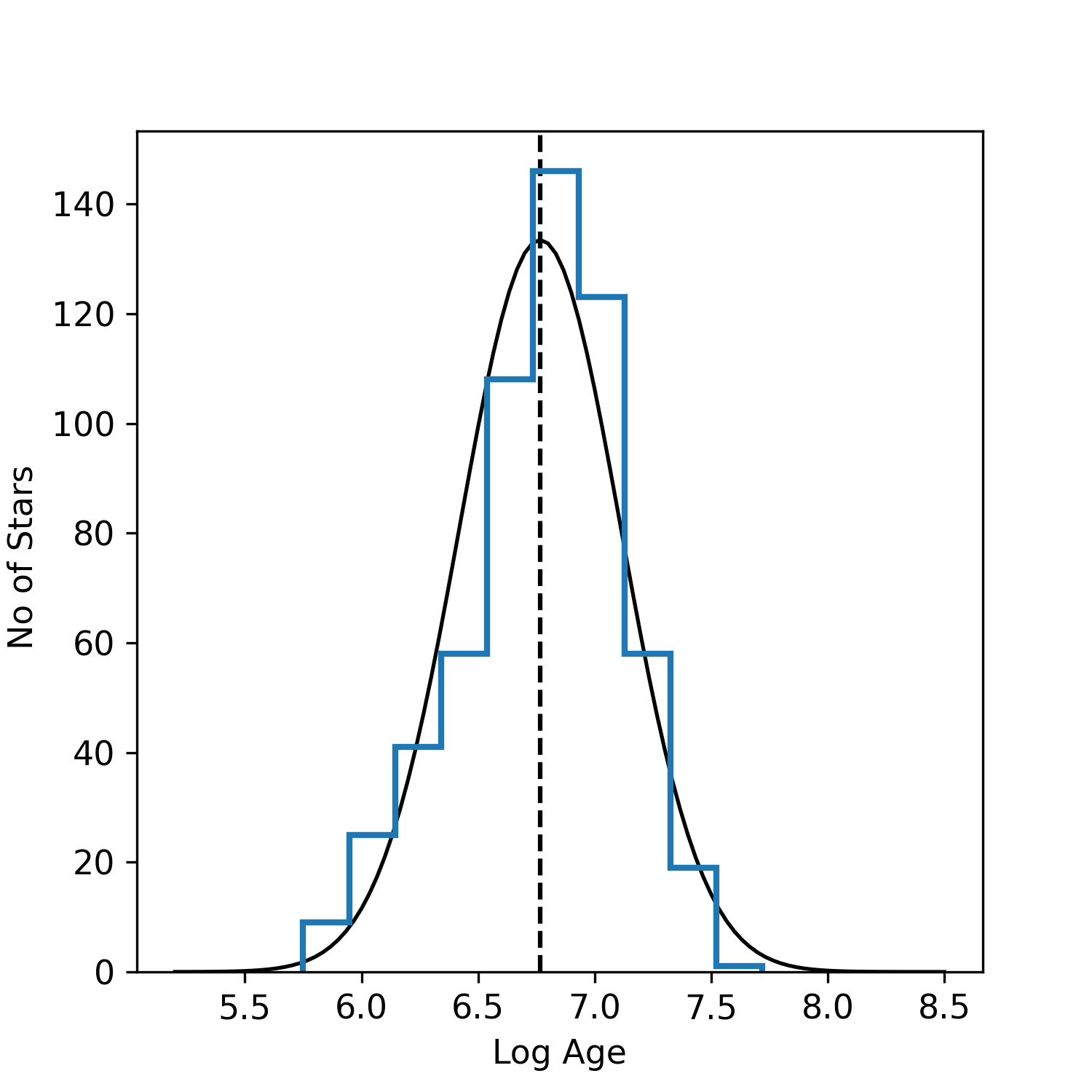}

\caption{Age distribution of the candidate members in the cluster Collinder 69 obtained from SED and HR diagram analysis. A Gaussian profile is fitted to the distribution, and the mean of the Gaussian is considered as the age of the cluster (black dotted line). 
}
\label{fig:c69age}
\end{figure}

 We determined the average age of the clusters by fitting a Gaussian distribution to the set of ages of its members (See \cite{2021MNRAS.504.2557D,2024ApJ...970...88P} for details). Figure \ref{fig:c69age} shows the age distribution of the sample cluster Collinder 69, fitted with the Gaussian curve.
After calculating the dataset's mean and standard deviation, we used the three-sigma iteration procedure to remove outliers that were greater than three-sigma. This procedure continued until the dataset showed no further changes (see \cite{2021MNRAS.504.2557D} for details). The mean of the Gaussian fit to this refined dataset represents the cluster's final age, and the standard error in the mean is taken as the corresponding uncertainty. The estimated cluster age values are displayed in Table \ref{table:main}. We compared our age analysis with the literature values for a few known clusters and found the results consistent. Our analysis estimates the age of IC 348 to be $\sim$ 3 Myr, which is consistent with estimates of 2-4 Myr found in the literature (\citealt{2018MNRAS.478.3674R} and references therein). For Collinder 69, we determined the age to be $\sim$ 6 Myr, which is in agreement with previous estimates of $\sim$ 5–6 Myr \citep{2009ApJ...707..705H,2011A&A...536A..63B}. For the relatively older clusters, such as IC 2602, Trumpler 10, and Melotte 22, we estimated average ages of $\sim$ 54 Myr, $\sim$ 27 Myr, and  $\sim$ 93 Myr, respectively, and the literature values are $\sim$ 50 Myr, $\sim$ 30 Myr, and  $\sim$ 100 Myr (\citealt{2022AJ....163..278N,2023A&A...675A.167P,2024A&A...689A..18A}). 

The observed luminosity dispersion in HR diagrams of young clusters, which often exhibit an order of magnitude spread at a given temperature, could be interpreted as a real age spread or as the result of a variety of observational and astrophysical variables. These aspects include stellar binarity, ongoing accretion processes, photometric variability, circumstellar disk evolution, and intrinsic uncertainties in evolutionary models \citep{2014prpl.conf..219S,2017ApJ...836...98J}. While we acknowledge these limitations, we believe that our methodology produces the most accurate age estimates possible with the photometric data available. Importantly, our uniform method for age determination ensures that any systematic biases are identical across all clusters in our sample. The relatively small range of Av values seen in our sample suggests that differential extinction will have little effect on our age estimates. Future spectroscopic identification of cluster members will definitely improve these age estimates and provide more accurate cluster parameters.

\section{Disk Fractions}
\label{sec:df}

The Disk Fraction is the ratio of sources with circumstellar disks around them to the total number
of sources within a cluster or star-forming region. Disk fraction analysis, which usually drops with age, shows how quickly disks degrade over time. Recent research has demonstrated that disk lifetimes and fractions can be metallicity-dependent \citep{2021AJ....161..139Y,2023A&A...675A.204V,2024ApJ...970...88P}. However, because our sample clusters are located in the solar neighbourhood, we consider solar metallicity a reasonable assumption for our study. Based on near-infrared excess, disk fractions of young clusters ($\sim$ 1-2 Myr) typically range from 80-90 \%, declining to 10-20 \% by $\sim$ 5-10 Myr, and less than 5 \% by $\sim$ 10-20 Myr \citep{2001ApJ...553L.153H,2007ApJ...662.1067H,2009AIPC.1158....3M,2021A&A...650A.157G}. Importantly, disk fractions have a considerable wavelength dependence, with longer wavelengths generally exhibiting greater fractions \citep{ribas2014}. Mid-IR observations (8-24$\mu$m) consistently indicate greater disk percentages than shorter wavelengths (1.2-4.6$\mu$m), reflecting the distinct disk areas probed by different wavelengths. In order to identify disk excess sources in various wavelength regimes, we used three different methods, each of which probed a distinct disk region. It is important to understand the completeness of our data sets and membership analysis before proceeding with the disk fraction analysis. Below, we discuss the completeness limits of our membership analysis, followed by the disk fraction analysis. 

\subsection{Completeness Limit}
\label{sec:massd}
\begin{figure*}
\centering
\includegraphics[width=2\columnwidth]{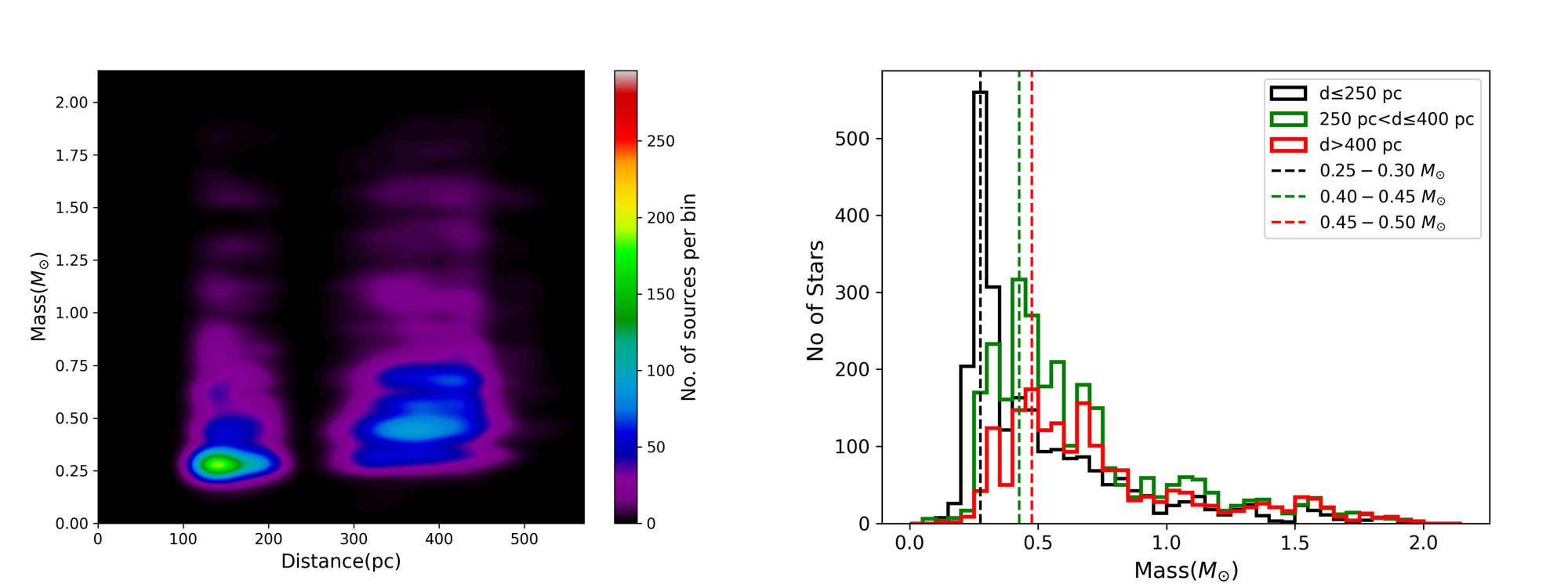}
\caption{Left Panel: Number of sources as a function of mass and distance. The color bar indicates the number of sources in each bin. Right Panel: Histograms showing the mass distributions; black indicates the sources within 250pc, green indicates sources with 250<d$\leq$400 pc and red indicates sources with d>400 pc. The dashed lines correspond to the peak bin in each case.
}
\label{fig:cdm}
\end{figure*}

The sample’s completeness level is crucial for determining the significance of
any statistical conclusion, as an incomplete sample would provide biased results. At a given mass, disk-bearing stars are brighter in the infrared than disk-less objects. Thus, depending on the distance and the stellar mass, we can lose some disk-less members in a given stellar cluster below the detecting limit of the infrared survey,  potentially resulting in inflated disk fractions. Hence, we checked for any biasing (in mass) corresponding to the distance of the source.

Figure \ref{fig:cdm} shows the mass (obtained from Section \ref{sec:agemass}) distribution of all the candidate members in this study as a function of their distance. The color bar indicates the number of sources. The figure shows two groups of stellar distribution with a gap of sources around 250 pc. We note that this gap is coming from the lack of target clusters in our list around this distance range, and it was a natural bias that occurred after implying the selection criteria (such as those regions with 100 sources and above; see Section \ref{sec:data} for details). The mass distribution of sources shows a bias as the distance increases. Below 250 pc, the peak density of sources is mostly in the mass range of $\sim$ 0.2 - 0.3\(M_\odot\), which is in good accordance with the peak mass of IMF in general \citep{2001MNRAS.322..231K,2002Sci...295...82K,2003PASP..115..763C} as well as with IMF of young clusters \citep[e.g.,][]{2021MNRAS.504.2557D}.
However, the peak mass range shifts to higher mass as the distance increases ($>$ 250 pc). The right panel of Figure \ref{fig:cdm} 
shows the histogram for mass distributions of the sources in different distance limits. For those sources within 250 pc, the mass peaks at $\sim$ 0.3 $M\odot$,  and for those within the distance limit of  250pc$<$d$\leq$400pc, the peak shifts to 0.4\(M_\odot\), whereas for d$>$400pc, the peak is further shifted to 0.45\(M_\odot\). Thus, for disk fraction analysis, we imposed a cut-off of sources above their mass completeness limits at the respective distance, i.e.,  all the candidate members within d$\leq$250pc are considered; for those within the distance range of 250pc<d $\leq$400pc, sources with mass more than 0.4\(M_\odot\) are included; and for clusters of d$>$400pc, sources with mass greater than 0.45\(M_\odot\) are included for the follow-up analysis. Since the $A_V$ associated with most of the clusters are < 1 mag (Table \ref{table:main}), we do not need to consider any effect on the completeness limit due to non-uniform reddening within the clusters.  The clusters under this study are not very rich in their stellar density. Hence, we need not consider the completeness limit due to the crowding effect.

\subsection{Disk fraction from J, H, K Band Analysis}
\label{sec:jhkex}
\begin{figure}
\includegraphics[width=1\columnwidth]{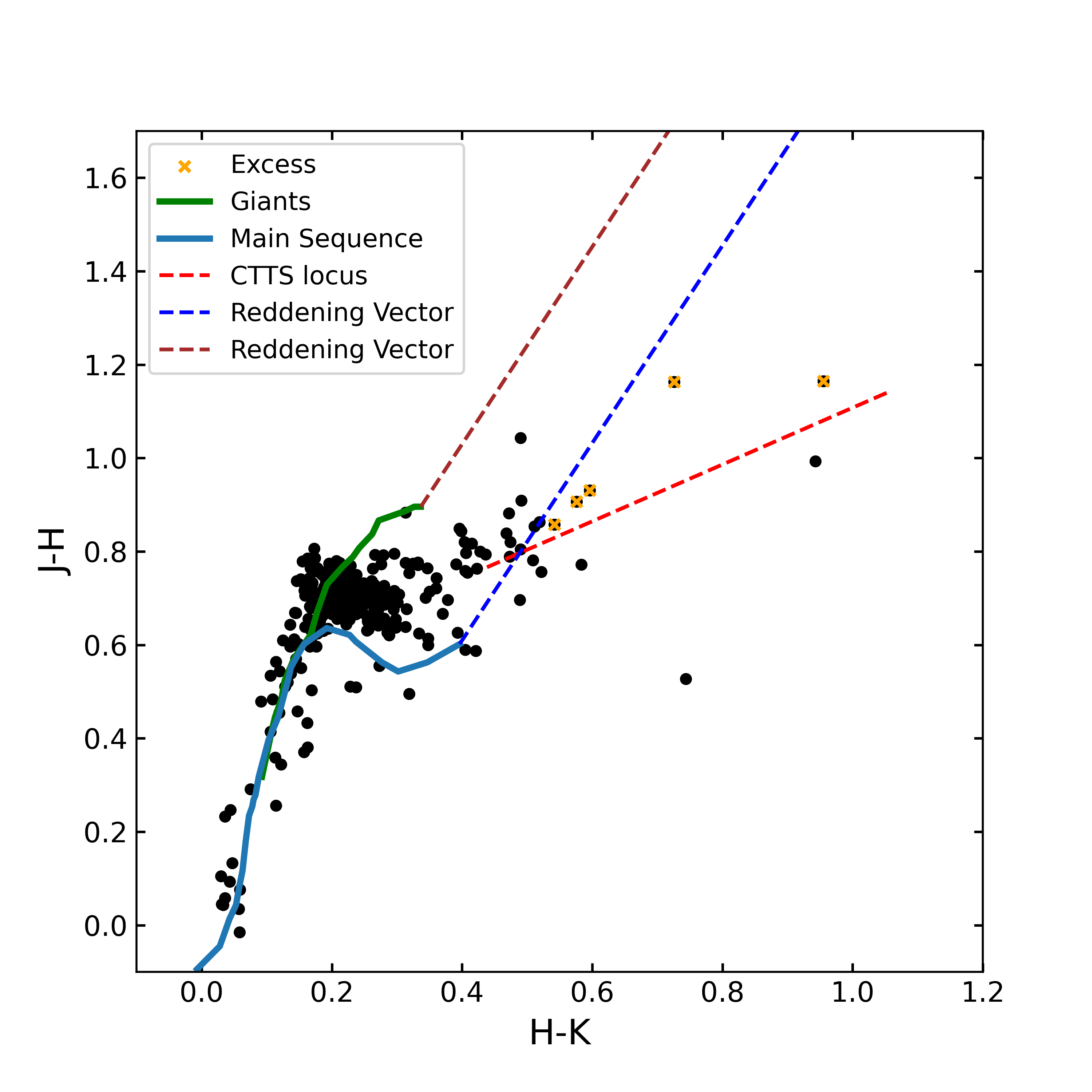}
\caption{J-H vs H-K CCD for the sample cluster Collinder 69. The green and blue curves represent the locus of giants and main sequence stars, respectively \protect\citep{1988PASP..100.1134B}. The red dashed line indicates the CTTS locus \protect\citep{1997AJ....114..288M} and the parallel dashed lines (blue and brown) indicate the reddening vectors (using the extinction law from \protect\cite{2019ApJ...877..116W}). Orange crosses are the sources with NIR excess due to disk around them. 
 }\label{fig:jhkccd}
\end{figure}

The J, H, and K bands (1.2-2.2µm)  pertain to the near-infrared (NIR) region, where excess in these bands signals the presence of an inner disk region. We employed J-H vs. H-K color-color diagrams (CCDs) to identify
stars with disks, indicating a greater H-K excess than diskless stars due to the IR dust emission. This has been a standard method for identifying the NIR excess sources in several studies (eg. \citealt{2001ApJ...553L.153H,2009ApJ...707..705H,2012MNRAS.424.2486J,2024ApJ...970...88P}).  The CCD for the sample cluster Collinder 69 is depicted in Figure \ref{fig:jhkccd} as an example. The red dashed line is the classical T Tauri star (CTTS) locus from \citet{1997AJ....114..288M}. The green and blue curves indicate the loci of giant and main sequence stars, respectively, as determined by \citet{1988PASP..100.1134B}. Two parallel dashed lines (blue and brown) represent the reddening bands from the tip of the main sequence and giant star sequence following the \citet{2019ApJ...877..116W} extinction law. The blue dashed line represents the boundary between the disk and diskless stars for objects in the low mass limit and is parallel to the reddening vector (see \cite{2024ApJ...970...88P} for details). All curves and lines in the figure are transformed to the 2MASS system using equations from \citet{2001AJ....121.2851C}.

The sources situated rightward of the reddening vector (blue dashed line in Figure  \ref{fig:jhkccd})  and above the CTTS loci, after accounting for photometric uncertainties, are considered stars with infrared excesses caused by circumstellar disks \citep{2002ApJ...565L..25S,Ojha_2004,2012ApJ...755...20S,2015MNRAS.454.3597D}.  Despite applying the correction for reddening, these stars do not return to the main-sequence locus owing to their significant near-IR excesses. On the other hand, those on the leftward of the reddening vector will move down along the slope towards the lower left after being adjusted for reddening, eventually moving to the group made up of main sequence stars. Some of these stars could have modest near-IR excesses. However, without spectroscopic information to estimate the reddening level,  it is impossible to confirm the presence of circumstellar disks. Thus, this work considers that these stars do not have near IR excesses.

The disk fraction for each cluster was determined by calculating the percentage of IR excess sources relative to the total number of sources within the completeness limit (Section \ref{sec:massd}). The Poisson error is considered as the
corresponding uncertainty in disk fraction estimation. The disk fraction estimated in J, H, K bands and the total number of candidate members in each cluster is listed in Table \ref{table:main}. Figure \ref{fig:cdf}, top left panel, shows the J, H, K disk fractions as a function of the age of the clusters. As the figure and table indicate, the fractions of sources with disk emission in  J, H, K bands are very minimal.  Our analysis confirms that the inner disks are depleted within 10 Myr, which is in agreement with the previous studies \citep{2001ApJ...553L.153H,10.1093/mnras/sty949,2024ApJ...970...88P}. The disk fraction values are very low since our sample consists of relatively evolved clusters ($>$ 3 Myr). 

\subsection{Disk fraction from H, K, W1 and W2 Band Analysis}
\label{sec:sec4.3}

The H, K, W1, and W2 bands correspond to the wavelength regime of 1.6 - 4.6 $\mu$m. \cite{2014ApJ...791..131K} have classified the YSOs into Class I, Class II, and transition disks based on 2MASS and WISE photometry. We used this classification scheme based on the H, K bands from 2MASS and WISE W1, W2 bands to identify class II sources from our sample. Since the clusters are relatively older, we do not have Class I sample in the list. 
The selection criteria for the respective bands were employed, and a source was identified as a class II source if the following conditions were satisfied:

\begin{equation}
\begin{aligned}\label{eq:first}
H - K & > 0.0 \\
H - K & > -1.76 \times (W1 - W2) + 0.9 \\
H - K & < (0.55/0.16)\times (W1 - W2) - 0.85 \\
W1 & \leq 13.0
\end{aligned}
\end{equation}

This classification scheme was applied to candidate members in each cluster, and the corresponding disk fraction was determined.  W3 and W4 band-based classifications were not employed, as these bands have fewer detections than the shorter wavelength bands (see section \ref{sec:w3,w4}). So, we analyzed them separately in the below section. 
The disk fraction estimated in H, K, W1 and W2 bands and the total number of candidate members in each cluster are listed in Table \ref{table:main}. The values of disk fraction are relatively higher compared to those of J, H, K-band analysis. 
Figure \ref{fig:cdf}, top right panel, displays the disk fraction as a function of the age of the cluster estimated using H, K, W1 and W2-bands. The disk fraction decays as the age increases and diminishes to zero after approximately 20 Myr. 

\subsection{Disk fraction from W3 and W4 Band Analysis}
\label{sec:w3,w4}

Compared to the W1 and W2 bands, the W3 and W4 bands have less sensitivity \citep{2014ApJ...791..131K,2016AJ....151...36L} and hence we have less detection in these bands. We employed a different scheme to identify the IR excess sources in these bands. \citet{ribas2014} developed a method to identify excess in a certain band based on the difference between observed and model flux. We employed a similar scheme where excess emissions were determined using a significant index \(\chi = \frac{{F_{\text{observed}} - F_{\text{model}}}}{{\sigma_{\text{observed}}}}\), where \(F_{\text{observed}}\) represents the observed flux, \(F_{\text{model}}\) corresponds to the respective model photospheric flux (obtained from SED fitting performed by VOSA as explained in section \ref{sec:param}), and \(\sigma_{\text{observed}}\) denotes the error at the corresponding wavelength. Excess is defined as a measurement with \(\chi \geq 5\), indicating a 5\(\sigma\) detection over the expected photospheric value.

\begin{figure}
\includegraphics[width=1\columnwidth]{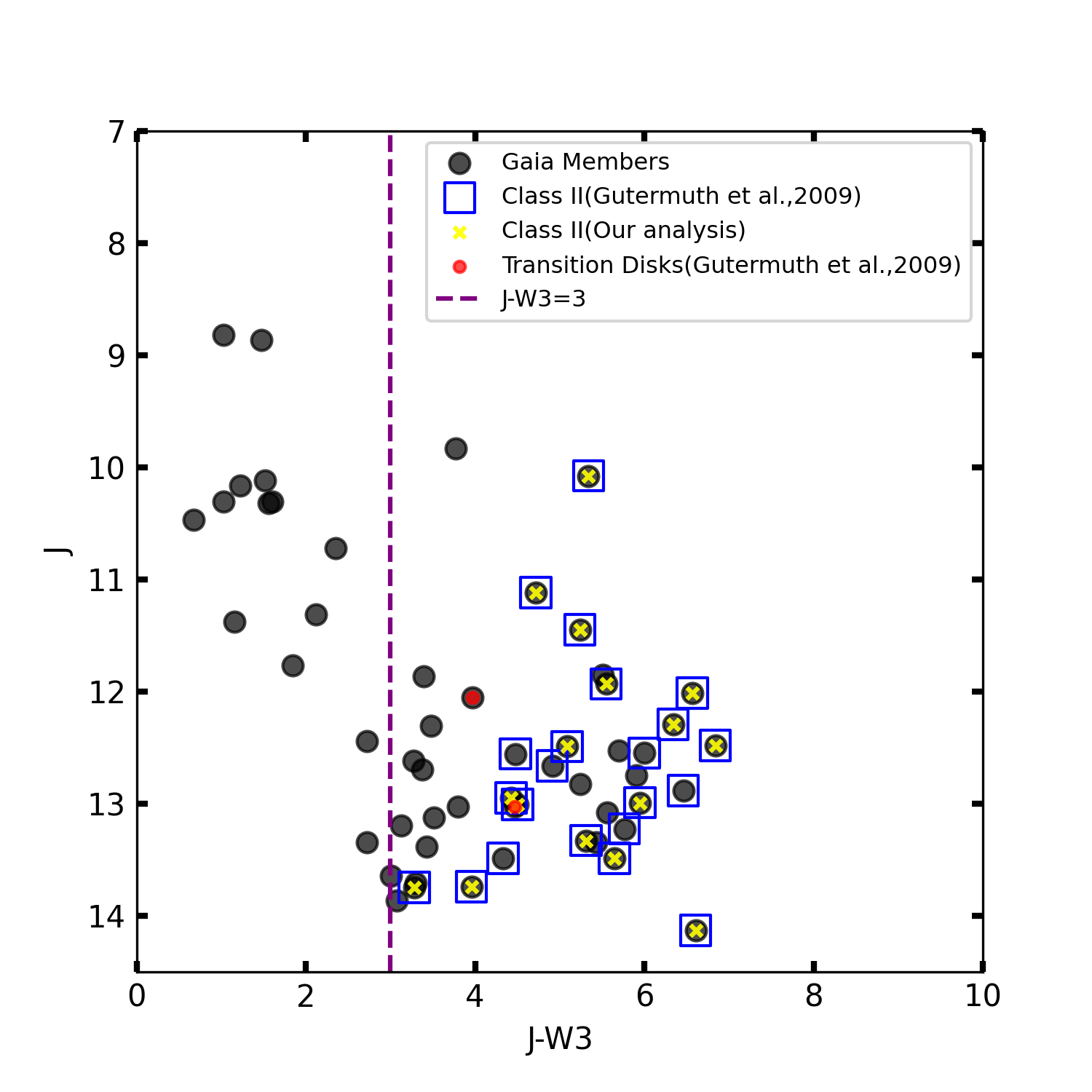}
\caption{J vs J-W3 CMD of a sample cluster IC 348. The black filled circles and yellow crosses represent the members considered and class II sources identified in the region from H, K, W1 and W2 bands. The red filled circles and blue boxes indicate the excess sources from \protect\cite{Gutermuth_2009}. The purple dashed line corresponds to J-W3=3 mag (see the text for details).}\label{fig:j_w3} 
\end{figure}
\begin{table*}
	
	\caption{Age and Disk Fraction of the Clusters across the different wavelength regimes.  The coordinates, distance and Av values are from CG2020. The square brackets indicate the explicit disk fraction.}
	\label{table:main}
	\begin{tabular}{llllllllll} 
		\hline
		Cluster & RA (deg)&Dec (deg)&Log Age &Distance(pc)&Av(mag) & DF\textsubscript{JHK}(\%) & DF\textsubscript{HKW1W2}(\%)& DF\textsubscript{W3}(\%) & DF\textsubscript{W4}(\%)\\
		\hline
		IC 348 & 56.132	&32.159&6.43±0.04&325±16&1.91&2±2 [1/51] &12±5 [6/51] &55±17 [17/31]&-\\ 
NGC 1980 & 83.810&-5.924&6.69±0.03&387±9&0.70&1±1 [1/76]&8±3 [6/74]&51±14 [21/41]&-\\
UBC 17a &83.422&	-1.671&6.74±0.03&363±6&0.80&1±1 [2/146]&4±2 [6/146]&16±4 [16/95]&-\\
Alessi 20&2.593	&58.742&6.77±0.03&433±13&0.60&0 [0/67]&10±4 [6/62]&22±7 [10/46]&-\\
Collinder 69&83.792	&9.813&6.77±0.01&406±20&0.25&1±1 [5/361]&12±2 [42/359]&30±4 [65/216]&-\\
UBC 17b&83.195	&-1.585&6.81±0.03&421±8&0.05&1±1 [1/74]&3±2 [2/73]&19±8 [7/37]&        -\\
Gulliver 6&83.278&-1.652&6.90±0.02&	422±19&0.25&0.5±0.5 [1/197]&2±1 [4/196]&12±4 [11/93]&-\\
ASCC 19&81.982&-1.987&6.96±0.02&361±12&0.13&0 [0/123]&2±1 [3/123]&11±4 [8/70]&-\\
ASCC 16&81.198&1.655&7.02±0.02&352±13&0.20&0.7±0.7 [1/148]&3±1 [4/147]&6±3 [4/66]&-\\
Pozzo 1&122.374	&-47.335&7.10±0.02&351±13&0.07&0 [0/218]&0 [0/213]&2±1 [3/140]&-\\
ASCC 127&347.205&64.974&7.16±0.02&380±12&0.43&0 [0/80]&1±1 [1/79]&2±2 [1/41]&-\\
RSG 8&344.983&59.371&7.18±0.03&452±42&0.53&0 [0/117]&0.8±0.8 [1/117]&19±7 [8/42]&-\\
UPK 422&88.007	&-7.543&7.23±0.02&302±11&0.26&0 [0/73]&0 [0/73]&0 [0/42]&-\\
RSG 7&344.190&59.363&7.28±0.04&428±17&0.57&0 [0/84]&0 [0/85]&19±8 [8/41]&-\\
NGC 2232&96.888&-4.749&7.32±0.02&326±11&0.01&0 [0/111]&0 [0/111]&0 [0/50]&-\\
NGC 3228&155.378&-51.814&7.33±0.03&491±18&0.08&0 [0/76]&0 [0/75]&2±2 [1/41]&-\\
UPK 640&250.260&-39.494&7.33±0.01&176±6&0.40&0 [0/394]&2±1 [7/392]&6±1 [18/301]&43	±13 [16/37]\\
BH 164&222.311	&-66.465&7.37±0.02&423±14&0.17&0 [0/115]&0 [0/115]&2±2 [1/43]&-\\
IC 4665&266.554	&5.615&7.37±0.02&349±16&0.45&0 [0/104]&0 [0/104]&0 [0/50]&-\\
NGC 2547&122.525&	-49.198&7.40±0.02&392±11&0.14&0 [0/153]&0 [0/153]&1±1 [1/87]&-\\
Collinder 140&110.882	&-31.966&7.41±0.02&386±16&0.00&0 [0/86]&0 [0/85]&0 [0/36]&-\\
NGC 2451B&116.128	&-37.954&7.42±0.02&368±13&0.18&0 [0/183]&0 [0/180]&0 [0/86]&-\\
Trumpler 10&131.943&	-42.566&7.44±0.01&437±14&0.00&0 [0/306]&0 [0/302]&14±4 [16/117]&-\\
RSG 5&303.482&	45.574&7.47±0.02&340±12&0.01&0 [0/108]&0 [0/108]&12±5 [7/57]&-\\
Collinder 135&109.362	&-37.044&7.48±0.01&305±11&0.01&0 [0/202]&0 [0/201]&0 [0/117]&-\\
UPK 535&127.025	&-50.899&7.49±0.03&329±10&0.07&0 [0/55]&0 [0/55]&3±3 [1/35]&-\\
Platais 8&136.718&	-58.685&7.62±0.02&135±5&0.14&0 [0/179]&0 [0/177]&2±1[3/148]&16±10 [3/19]\\
IC 2391&130.292	&-52.991&7.70±0.02&152±3&0.04&0 [0/189]&0 [0/185]&0.7±0.7 [1/135]&0 [0/14]\\
NGC 2451A&115.736	&-38.264&7.70±0.02&194±4&0.00&0 [0/281]&0 [0/277]&0 [0/139]&-\\
IC 2602&160.613	&-64.426&7.73±0.02&152±4&0.03&0 [0/216]&0 [0/208]&0 [0/179]&17±9 [4/24]\\
Platais 9&139.241&-43.862&7.74±0.03&183±8&0.00&0 [0/99]&0 [0/98]&2±2 [1/57]&-\\
Melotte 22&56.601&24.114&7.97±0.02&136±3&0.18	&0 [0/878]&0 [0/875]&2±1 [8/422]&16±6 [8/49]\\
		\hline
	\end{tabular}
\end{table*}

\begin{figure*}
\includegraphics[width=2\columnwidth]{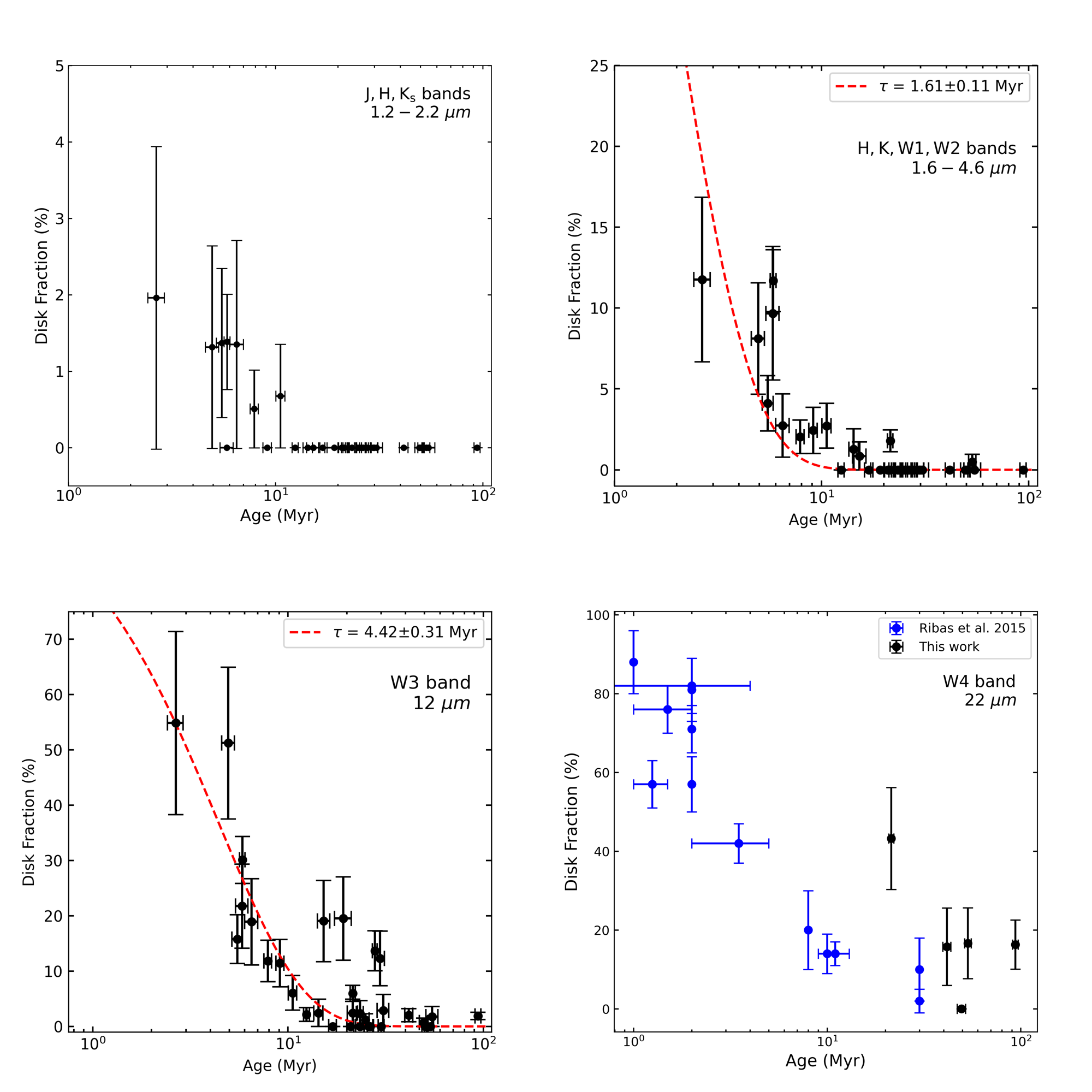}
\caption{Top left (Panel 1), Top right (Panel 2), Bottom left (Panel 3) and Bottom right (Panel 4) displays the disk fractions for all regions across (J,H,K), (H,K,W1,W2), W3 \& W4 bands. The red dashed curve in panel 2 and panel 3 represents the exponential fit to the dataset as explained in Section \ref{sec:dis1}}\label{fig:cdf} 
\end{figure*}

We also employed additional color cut-offs for W3 and W4 bands to refine our selection. For the W3 band, we used the distribution of known YSOs of the well studied cluster in the list, i.e., IC 348. Figure \ref{fig:j_w3} displays the J vs J-W3 CMD for the cluster IC 348. We selected excess sources from  \cite{Gutermuth_2009} based on Spitzer data (indicated by red filled circles and blue squares). We included the  candidate members in this study  (black filled circles) and the class II (yellow crosses) sources identified using the criteria in Section \ref{sec:sec4.3}. 
The figure shows that most of the excess sources from the literature and class II sources from our analysis lie to the right of J-W3=3 mag, shown as the purple dashed line.  Thus, we chose an additional color cut off of \(J-W3 \geq 3\) mag to select the  excess sources in the W3 band. Thus, W3 band excess sources are those that satisfied both conditions: \(\chi \geq 5\) and \(J-W3 \geq 3\) mag. 
In the case of the W4 band, the colour cut-off of \(K-W4 \geq 3.55\) mag based on the criteria given in \cite{2014ApJ...784..126E} is applied. Thus, W4 excess sources are those with \(\chi \geq 5\) and \(K-W4 \geq 3.55\) mag. As mentioned in section \ref{sec:massd}, since the dust is optically thin for larger wavelengths, the extinction affects less the radiation in the mid-infrared range (e.g., W3, W4 bands). Since there is limited detection of sources as the distance of the cluster increases, we retained only those clusters within 250 pc distance for the disk fraction analysis using the W4 band. This is because, for the distant clusters ($>$ 250 pc), the detections in the W4 band could be biased towards the excess sources since they appear brighter in the W4 band. Thus, these clusters appeared to have very high disk fraction values. Additionally, we considered only those clusters with a number of detections greater than 10 in the W4 band \citep{ribas2014}. After implementing all the above conditions, we retained only 5 clusters for disk fraction analysis in the W4 band.  To improve statistics, we included additional clusters from the list of \citep{2015A&A...576A..52R} [hereafter "RBS2015"] in our W4 analysis, as we follow the same methodology. In summary,  the disk fractions were calculated for each cluster in the W3 and W4 bands based on the above criteria and are listed in Table \ref{table:main}. The respective disk fractions as a function of age in the W3 and W4 bands are shown in Fig \ref{fig:cdf}, bottom left and right panel, respectively. The additional sources from RBS2015 are shown as blue dots in the bottom right panel of Fig \ref{fig:cdf}.  The disk fraction values are higher at longer wavelengths than at shorter ones, with $\approx 20\% $ for older clusters ($> 20$ Myr). These long-lived disks are essential for understanding disk evolution, which is detailed in section \ref{sec:dis1}.

\subsection{Comparison of disk fraction with previous studies}

Further analysis compared the disk fraction (DF) values reported by RBS2015 for the IC 348 and Collinder 69 clusters with our findings. It is noted that our methodology differs from RBS2015
in the short-wavelength analysis. As explained in Section \ref{sec:sec4.3}, we implemented Koenig’s scheme, although  RBS2015 took a different approach (described in Section \ref{sec:w3,w4}).  For IC 348 and Collinder 69, they reported DF\textsubscript{short} values of 17±2 and 10±4, respectively. Despite the methodological variations, these results are in agreement with our DF\textsubscript{HKW1W2} findings of 12±5 and 12±2.
In the intermediate wavelength (8-12 $\mu$m) range, RBS2015 reported DF\textsubscript{IM} (DF in intermediate wavelength range) values of 41±3 for IC 348 and 23±6 for Collinder 69, agreeing with our DF\textsubscript{W3} values of 55±1 and 30±4, respectively. The DF\textsubscript{IM} values are slightly lower because they are based on the IRAC 8  $\mu$m band, whereas our DF\textsubscript{W3} values are based on the 12  $\mu$m band. 
Disk emission is usually greater at 12 $\mu$m than at 8 $\mu$m because of the disks's spectral energy distribution usually tends to rise toward longer wavelengths in this region. 
Consequently, our analysis at 12 $\mu$m is more sensitive to the presence of disks, particularly those with weaker or evolved emissions, leading to slightly higher disk fraction values.

\subsection{Age-Binned Analysis}
\label{sec:age bin}
In the cluster-based disk fraction analysis, we had very few detections within individual clusters in the W3 and W4 bands. We excluded those clusters with detections less than 10 in the W4 band from the disk fraction analysis, as explained in Section \ref{sec:w3,w4}. As a result, we had a limited number of clusters with W4 excess.  One way to mitigate this problem is to divide our sample into various age bins. This approach allows us to include those sources in the clusters with fewer than 10 detections in the W4 band. Combining them into various age bins gives us adequate statistical information to analyze it robustly  \citep{2020ApJ...893...56M}. 
\begin{table*}
	\centering
\caption{Disk Fraction across  wavelengths in different age bins}
\label{table_3}
\begin{tabular}{lllllll}
\hline
Bin&Range (Myr)&Median Age (Myr)&DF\textsubscript{HKW1W2}(\%) &DF\textsubscript{W3}(\%) & DF\textsubscript{W4}(\%)\\
\hline
1&1-6&6&10±1[660]&31±3[408]&92±11[132]\\
2&6-10&8&2±1[337]&13±3[174]&86±23[29]\\
3&10-17&12&0.9±0.4[629]&5±1[331]&44±12[43]\\
4&17-31&25&0.4±0.1[1902]&4±1[1032]&50±8[103]\\
5&42-60&50&0.1±0.1[945]&0.8±0.3[657]	&12±5[65]\\
6&90-100&94&0.0[875]	&2±1[422]	&16±6[49]\\
\hline
\end{tabular}

\end{table*}
Thus, we divided our sample into 6 bins across the age range of 1-100 Myr, employing a randomized bin selection process. In this analysis, we included all good detections from the W3 and W4 bands while also applying the mass-distance criteria outlined in Section \ref{sec:massd}. Each source was assigned its parent cluster's age for the follow-up analysis. For each bin, we calculated the median age of the stars within that bin. The age distribution was continuous up to 31 Myr, after which we observed two significant gaps where no clusters were present in our sample: between 31-42 Myr and 60-90 Myr. The analysis of disk fraction at different wavelength regimes follows the methods described in earlier sections. We excluded the J, H, and K-band based disk fraction analysis in this section because it is evident that DF\textsubscript{JHK} falls to zero beyond 10 Myr (Section \ref{sec:jhkex}).  The disk fraction across different wavelengths in 6 age bins is listed in Table \ref{table_3}. The disk fraction varies as a function of age and at different wavelengths.  We tested various age ranges for binning and found that all yielded similar disk dissipation trends across all wavelengths. The slopes in the logarithmic planes of disk fraction and age (see Section \ref{sec:dis} for details) are comparable regardless of the chosen age ranges.

\subsection{Disk Classification and Characteristics}
\label{dclass}
Circumstellar disks go through several evolutionary phases, which have been characterized using different classification techniques and names in the literature. For example, \cite{2012ApJ...747..103E} and \citet{2012ApJ...758...31L} provide such standard categorization for disks, as seen below. 
Full disks are optically thick at infrared wavelengths, retaining their primordial dust and gas without evident disk dissipation. Transitional disks exhibit small or no excess in the near-IR, still showing large IR excesses at larger wavelengths, suggesting the presence of inner holes larger than the dust sublimation region. In the pre-transitional disk phase, the disk still shows near-IR excesses, suggesting the presence of large gaps in the dust distribution. On the other hand, evolved disks exhibit an overall decrease in infrared excesses compared to the full disk, suggesting a more homogeneous radial evolution of the disk. Finally, debris disks, composed of second-generation dust created by planetesimal collisions, only show modest mid-IR excesses.

In our sample, we consider the excess sources from J, H, K bands (Section \ref{sec:jhkex}) and Class II sources (Section \ref{sec:sec4.3}) to be full disk sources since they display excess in near-IR wavelengths. We detected 94 full disk sources in total, ranging from $\approx$ 3 to 21 Myr,  representing the population of disk in which the dispersion processes have not significantly affected their primordial structure. While these sources primarily trace younger full disks, we note that older disk-bearing stars, which are discussed later (Section \ref{sec:olddisk}), may also host primordial disks despite not always displaying prominent near-IR excess. Transitional disks are particularly interesting because they represent a crucial evolutionary stage between primordial and debris disks. These objects have an optically thin inner region, which may have been cut out during planet formation, and an optically thick outer region. This unique structure produces a distinct spectral energy distribution: transitional disks often show little to no excess in the near-IR but display excess emission at mid-IR wavelengths comparable with full disks. Because of this unique signature, color-color diagrams (CCDs) combining near- and mid-IR bands, notably K-W3 versus K-W4, serve as useful tools \citep{2012ApJ...758...31L,2014ApJ...784..126E,2022AJ....163...74T,2023AJ....165..205H} for identifying transitional disks in our sample.

\begin{figure}
\includegraphics[width=1\columnwidth]{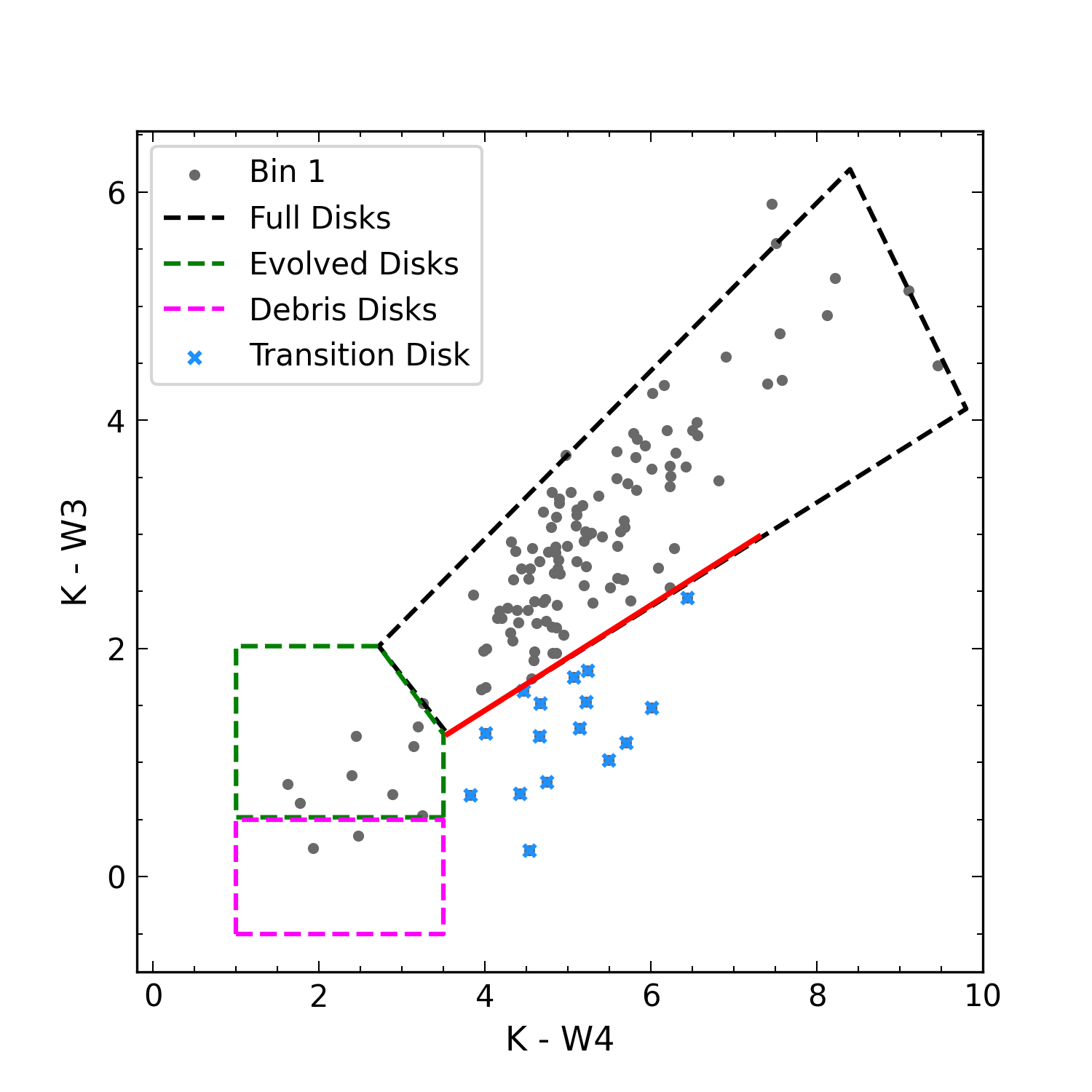}
\caption{K-W3 vs K-W4 CCD for Bin 1. The grey dots represent all the sources in the bin and the different regions corresponds to different types of disks. The blue crosses indicate the transition disks. The red line indicates the boundary between full disks and transitional disks. }\label{fig:td}
\end{figure}

We used this CCD approach on our age-binned data to examine if transitional disks were present in our sample. As an example of the procedure, Figure \ref{fig:td} illustrates the K-W3 vs K-W4 CCD for Bin 1. Based on criteria established in previous studies \citep{2007ApJ...662.1067H,2012ApJ...758...31L,2018AJ....156...75E}, we identified the regions corresponding to full disks, evolved disks, debris or evolved transitional disks, and transitional disks as shown in the plot. Specifically, objects that satisfied the two criteria of (K-W4) > 3.55 and falling below the line (K-W3) = 0.46*(K-W4) - 0.39 were identified as transitional disks.

These transitional disk candidates are marked with blue crosses
in the figure. This technique yielded 33 transitional disks across all six age groups. These sources have excess emissions in the W4 band, which is compatible with the expected transitional disk spectral features. This collection of 33 transitional disk candidates is an excellent resource for investigating their characteristics and evolutionary histories in future.

\section{Search for  Accretors}
\label{sec:accr}
Star formation and the features of planetary formation and migration zones are significantly shaped by the 
accretion process through protoplanetary disks \citep{2010A&A...510A..72F,2019AJ....157...85B,2022AJ....163...74T}. In low-mass stars, the magnetic field \citep{2016ARA&A..54..135H} drives accretion, which occurs when material falls from the inner circumstellar disk to the surface. Hot spots form on the stellar surface owing to material accretion, resulting in increased Balmer and other line emissions  \citep{1988PASP..100.1134B,Muzerolle_2003}. Current research suggests that the period of accretion is similar to the average lifespan of primordial disks, which is less than 10 Myr  \citep{2010A&A...510A..72F}. However, there are notable instances of accretion occurring at ages exceeding 10 Myr \citep{2018MNRAS.476.3290M}. Some examples are the 45 Myr old WISE J080822.18-644357.3 in the Carina association and the 42 Myr old 2MASS J0501-4337 in the TW Hydrae association \citep{Silverberg_2020}. These long-lived disks, which are frequently called "Peter Pan" disks, challenge our knowledge of disk evolution.  Since this study aims to understand the disk evolution of relatively older clusters,  we searched for any accretors within our sample utilizing the low-resolution spectra from the archive of LAMOST DR8 (Section \ref{sec:lamost}). 
We examined accretion signatures using LAMOST optical spectra from our sample of 773 sources.
These sources belong to clusters ranging from 1 to 30 Myr and one older cluster at 92 Myr.

One important tracer of accretion in YSOs is the H$\alpha$ emission line. We can determine the H$\alpha$ line luminosity (L\textsubscript{ H$ \alpha $}) using its equivalent width (EW). Using this luminosity, we can compute the accretion luminosity (L\textsubscript{acc}) and, hence, the mass accretion rate \citep{2017A&A...602A..33F,2017A&A...600A..20A,2023JApA...44...67A,2023JApA...44...75N}. We estimated the EW of the H$\alpha$ line using a Python package called PHEW  (PytHon Equivalent Widths, \cite{2022zndo...6422571N}).  PHEW fits a baseline to the spectrum's continuum using a polynomial of a certain order. The spectral feature of interest is then fitted with a Voigt profile, which isolates it from the continuum by excluding certain wavelength ranges surrounding the feature. The area under the fitted Voigt profile with respect to the baseline yields the EW. The uncertainties in the EW are obtained using the Monte Carlo (MC) method of error propagation, assuming Gaussian distributions for the uncertainties in spectral fluxes.

\begin{figure}
\includegraphics[width=1\linewidth]{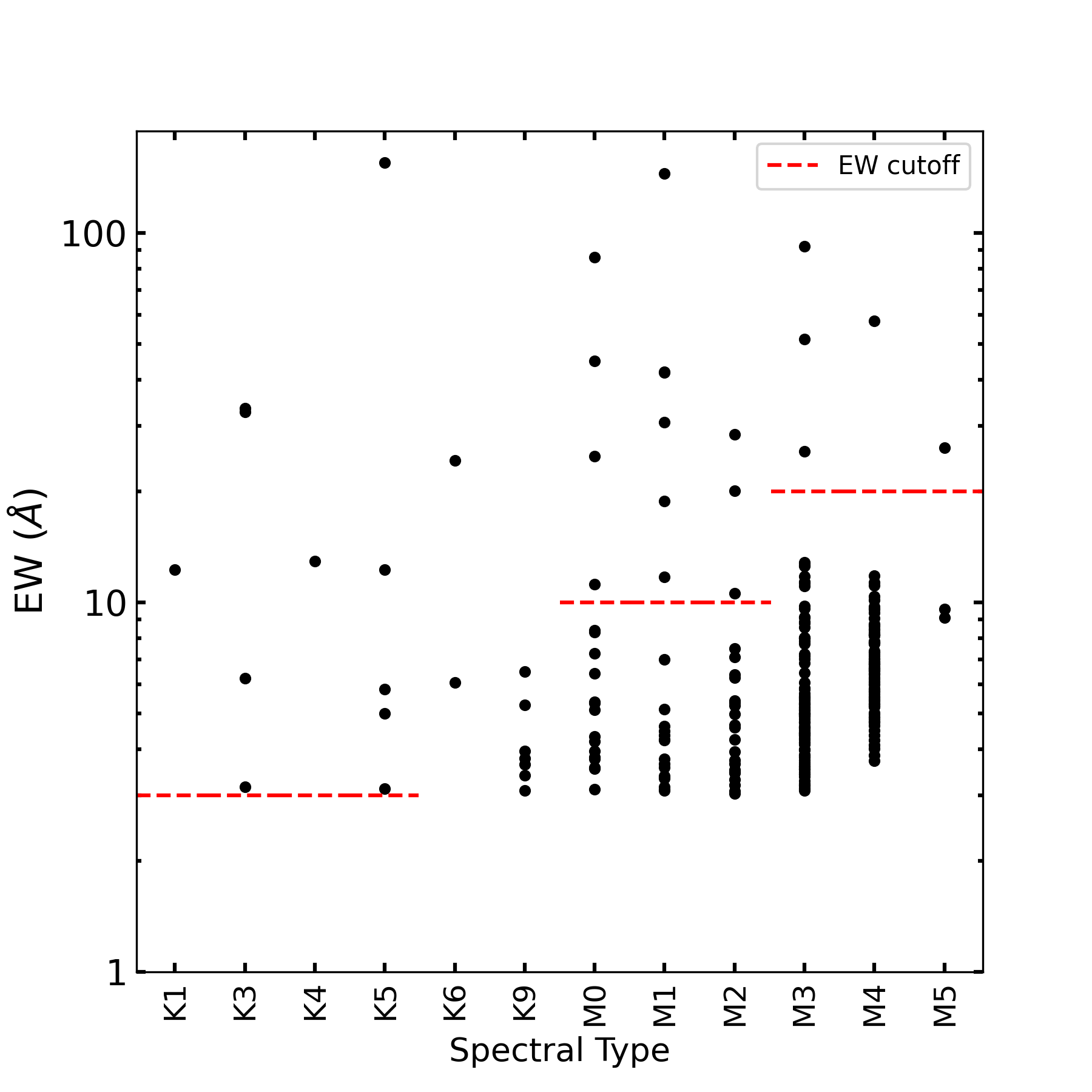}
\caption{EW(H$ \alpha $) vs Spectral Type plot. The red dashed line indicates the cut-off for accretors}\label{fig:figure2}
\end{figure}

Spectral types of the sources were assigned by comparing intrinsic and observed  temperature (T\textsubscript{eff}) values (which we obtained from SED fitting as explained in Section \ref{sec:param}) with Table 6 from \cite{Pecaut_2013}. We identified the accretors in our sample based on the EW(H$\alpha$) cut-offs for different spectral types from \cite{White_2003}, i.e.,  3\AA~ for K0–K5 stars, 10\AA~  for K7–M2.5, 20\AA~  for M3–M5.5 stars and 40\AA~  for M6–M7.5 stars. Figure \ref{fig:figure2} illustrates the EW vs. spectral type plot, where the red dashed line indicates the cut-offs. We identified 29 accretors in our sample.

Thus, we can finally determine the mass accretion rate of the 29 accretors. We begin with analysing continuum flux density at H$\alpha$ by using the standard relation,
\begin{equation}
\begin{aligned}\label{eq:2.1}
 F_{\text{R}}(H\alpha)= F_{\text{0}} \times 10^{-\frac{R_0}{2.5}}
\end{aligned}
\end{equation}
where \(F_{0}=2.8307 \times 10^{-12} \, \text{W} \, \text{m}^{-2} \, \text{A}^{-1}\) and \(R_{0}\) is the extinction corrected R band magnitude from Pan-STARRS1 
 \citep{2016arXiv161205560C}; we are taking this R-band flux density as a proxy for the continuum flux density underlying H$\alpha$ line \citep{Mathew_2018}.

The H$\alpha$ line flux (F\textsubscript{ H$ \alpha $}) can be determined from the relation,
\begin{equation}
\begin{aligned}\label{eq:2}
 F_{\text{H}\alpha} = \text{EW(H}\alpha) \times F_{\text{R}}(H\alpha)
\end{aligned}
\end{equation}
The L\textsubscript{ H$ \alpha $} is then calculated using,
\begin{equation}
\begin{aligned}\label{eq:3}
 L_{\text{H}\alpha} = 4\pi d^2 \times F_{\text{H}\alpha}
\end{aligned}
\end{equation}
where d is the distance of the source. \cite{2009A&A...504..461F} provides a relation between L\textsubscript{acc} and L\textsubscript{ H$ \alpha $} as follows,
\begin{equation}
\begin{aligned}\label{eq:4}
 \log(L_{\text{acc}})=(2.27 \pm 0.23) + (1.25 \pm 0.07) \times \log(L_{\text{H}\alpha}).
\end{aligned}
\end{equation}
The Mass accretion rate($\dot{M}_{\text{acc}}$) and L\textsubscript{acc} is related by the free fall equation:
\begin{equation}
\begin{aligned}\label{eq:5}
\dot{M}_{\text{acc}} = \frac{L_{\text{acc}} \cdot R_{*}}{G M_{*} \left(1 - \frac{R_{*}}{R_{\text{in}}}\right)}
\end{aligned}
\end{equation}
where M\textsubscript{*} and R\textsubscript{*} are the stellar mass and radius, respectively. G is the Gravitational Constant. Here we adopt R\textsubscript{in}=5R\textsubscript{*} \citep{1998ApJ...492..323G,2008ApJ...681..594H,Manara_2016}. R\textsubscript{*} is determined based on the temperature luminosity relation of the star. Thus, the mass accretion rates of the sources were calculated. We estimated mass accretion rates for 29 sources in the list, and their accretion rates lie in the range of $10^{-8} - 10^{-11} \, \mathrm{M}_\odot \, \text{yr}^{-1}$. These estimates are discussed in Section
\ref{sec:macc}. 
\section{Discussions}
\label{sec:dis}
Our research of disk-bearing sources in several clusters sheds light on the evolution of protoplanetary disks and their relationship to stellar characteristics. The wavelength dependence of disk fractions, the existence of older disk candidates, the correlation between disk longevity and stellar mass, and the accretion activity among our sources are all examined in this section. These factors lead to a greater understanding of disk evolution timescales across varying stellar masses.
\subsection{Wavelength Dependent Disk Fractions}
\label{sec:dis1}
It is well known that disk fraction decreases as cluster age increases, as demonstrated by previous studies \citep{2001ApJ...553L.153H,2007prpl.conf..573M,2007ApJ...662.1067H,ribas2014,Richert_2018,2022ApJ...939L..10P,2023JApA...44...77D,2024ApJ...970...88P}. Our analysis confirms that
the disk fraction decreases as systems age, consistent across all wavelengths.
Photoevaporation models indicate that disk dissipation timescales vary between the inner and outer disk regions \citep{2009ApJ...705.1237G}. 
 \cite{ribas2014} analyzed the dissipation timescales in the inner (0.01 - 1 AU) and intermediate disk regions (0.03- 5 AU), discovering that the timescales are longer at longer wavelengths. We also observe a similar slow decay as the wavelength increases. The excess in short wavelengths generally corresponds to the presence of primordial, optically thick disks 
  \citep{Williams_2011} characterized by disk
radii ranging from 0.01 to 1 AU. The top panels of
figure \ref{fig:cdf} confirm that these thick disks decay within the first 20 Myr.

\begin{figure*}
\includegraphics[width=1\linewidth]{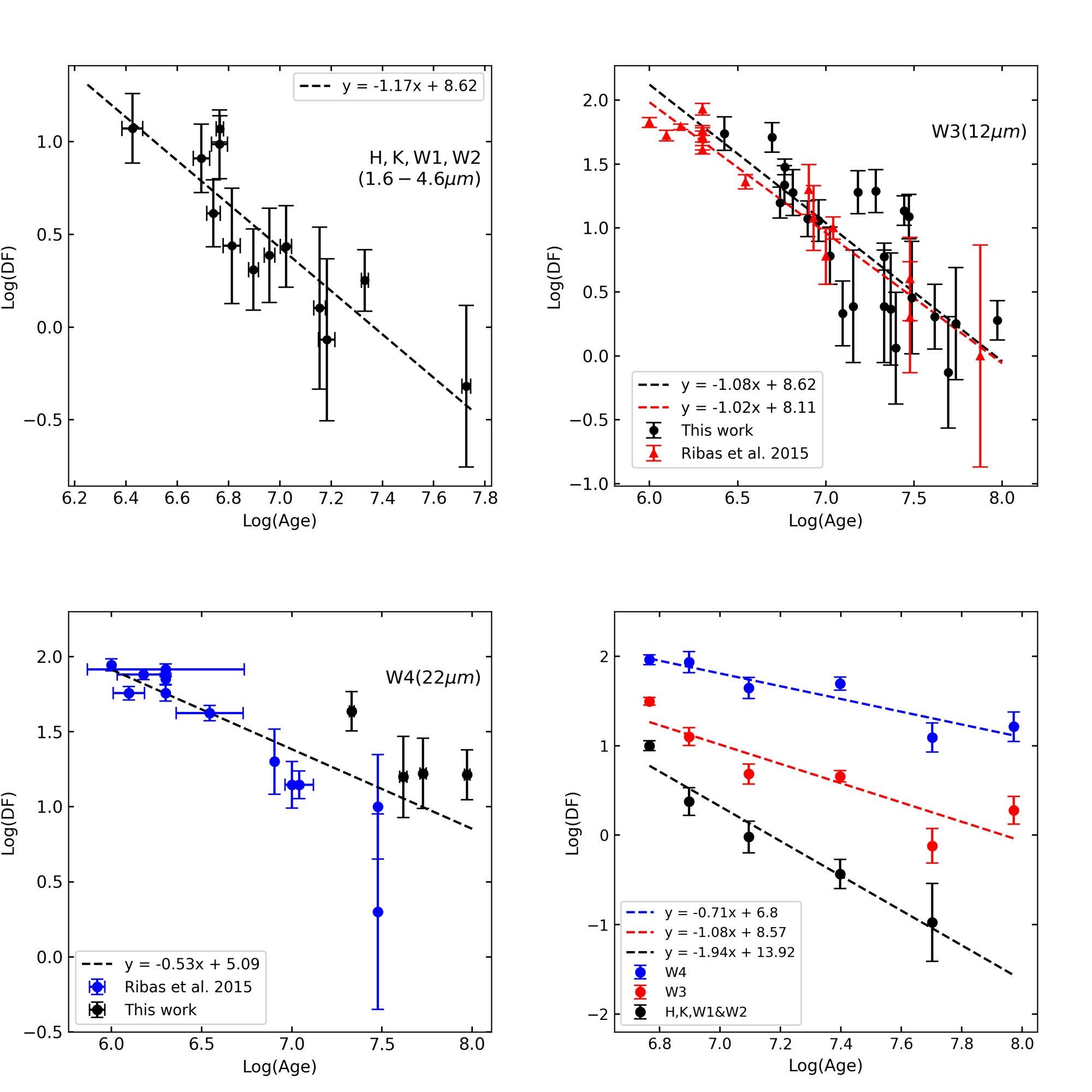}
\caption{Top left (Panel 1),Top right (Panel 2) and Bottom left (Panel 3) display the disk fractions for all regions across (H,K,W1,W2), W3 \& W4 bands in the logarithmic plane. The black dashed line represents the best-fit linear line. The red triangles in Panel 3 indicate the sources from RBS2015, and the red dashed line is the best fit for the sample. The bottom right (Panel 4) shows the disk fraction results of the age-binned analysis in the logarithmic plane. The blue, black \& red colors correspond to the wavelength regions (H,K,W1,W2), W3 \& W4 respectively. The coefficients of linear fits are mentioned in each panel. }\label{fig:log_plot}
\end{figure*}

Compared to the shorter wavelength bands, the W3 and W4 (bottom left and right panels in Figure \ref{fig:cdf}) analysis reveals a
more gradual decline in the disk fraction as a function of age, which indicates a prolonged disk lifetime in this regime. Also, the disk fraction values are higher in these regimes than in the short wavelengths, although the numbers of disk and diskless population detections are relatively smaller. One caveat to be noted is that the higher disk fraction values in this context could be attributed to the increased detection of excess sources in the W3 and W4 bands relative to the non-excess sources because of their enhanced brightness. Thus, the few clusters showing enhanced disk fractions in the W3 band at $\sim$15-20 Myr  (Figure \ref{fig:cdf} bottom left panel) must be taken cautiously.
High-sensitive deeper observations in MIR bands are essential for an unbiased sample. 
W3 corresponds to the disk region with radii ranging from 0.03 to 5 AU, while W4 corresponds to the disk region with radii ranging from 0.3 to 60 AU \citep{ribas2014}. 

The exponential decay model provides a standardized approach to quantify disk lifetimes across the entire time domain using a single function \citep{2009AIPC.1158....3M}. We fit our data with an exponential function of the form $f(t) = Ae^{-t/\tau}$, where $t$ is the cluster age (in Myr), and $\tau$ is the characteristic time. Since our sample lacks clusters younger than 2.6 Myr, we adopted $A = 100$ \citep{2010A&A...510A..72F,2019AJ....157...85B}  to properly anchor the initial disk fraction. This parameter $A$ quantifies the primordial disk population at birth (t=0), before significant evolution occurs.

Our characteristic timescales are $\tau_{\text{short}} = 1.6 \pm 0.1$ Myr for the shorter wavelength bands (H, K, W1, and W2; 1.6-4.6 µm) and $\tau_{\text{W3}} = 4.4 \pm 0.3$ Myr for the W3 band (12 µm) as shown in figure \ref{fig:cdf} top right and bottom left panels. The $\tau_{\text{short}}$ value is slightly lower than the typical range of 2-3 Myr reported in the literature \citep{2006ApJ...651L..49C,Williams_2011,ribas2014}, which may be attributed to our sample's lack of very young populations (<2.6 Myr) combined with our inclusion of numerous older regions (>10 Myr) where disk fractions approach zero.

In contrast, our $\tau_{\text{W3}}$ is higher than the 2.8 Myr timescale reported by Ribas et al. (2014), who also included 8um Spitzer photometry. In addition to the different datasets, the small differences between our estimations and Ribas's timescales are likely due to our detection of more excess sources in evolved clusters, indicating longer-lived disk components at this wavelength. When comparing $\tau_{\text{short}}$ and $\tau_{\text{W3}}$, the significantly longer timescale at 12 $\mu m$ confirms the inside-out clearing scenario of protoplanetary disk evolution, with the inner disk (traced by near-IR bands) dissipating more rapidly than the outer disk regions (traced by mid-IR bands). We did not fit an exponential function to the W4 (22 $\mu m$) data because this wavelength regime traces both primordial and second-generation debris disks, which follow different evolutionary pathways and cannot be accurately modeled with a single exponential decay function.

Another way to mathematically express the disk dissipation time scales is to put the respective disk fractions and age in the logarithmic plane \citep{2024ApJ...960...58H}. We fit a linear equation of the form y=mx+c onto the log(DF) and log(age) values. Figure \ref{fig:log_plot} displays these representations from top left to bottom right. The first three panels are the results of our cluster-based disk fraction analysis. The black dashed line and black dot represent the linear fit and the respective disk fractions in each panel. In panels 2 and 3, we have included the disk fractions from RBS2015 for intermediate (8-12 $\mu$m; red dots) and long (22-24 $\mu$m; blue dots) wavelengths. The red dashed line in panel 2 indicates the linear fit from the RBS2015 data. The respective equations are given by:
\begin{equation}
\log(DF_{\text{H,K,W1,W2}}) = (-1.17\pm0.19) \times \log(Age) + (8.62 \pm 1.30)
\end{equation}
\begin{equation}
\begin{aligned}\label{eq:logw3}
\log(DF_{\text{W3}})=(-1.08\pm0.19) \times \log(Age) + (8.62 \pm 1.37)
\end{aligned}
\end{equation}
For RBS2015,
\begin{equation}
\begin{aligned}\label{eq:logIM}
\log(DF_{\text{IM}})=(-1.02\pm0.06) \times \log(Age) + (8.11 \pm 0.43)
\end{aligned}
\end{equation}

where $\log(DF\textsubscript{IM})$ represents the disk fractions in intermediate wavelength regions in RBS2015.\\
For the combined sample of clusters in our analysis and from RBS2015, in long wavelength (W4 band),
\begin{equation}
\begin{aligned}\label{eq:logw4}
\log(DF_{\text{W4}})=(-0.53\pm0.08) \times \log(Age) + (5.09\pm 0.51)
\end{aligned}
\end{equation}

The slopes of the linear fit indicate the disk decay rate of the sample. As we move on from (H, K, W1, W2) to W4 bands, the slope decreases consistently, with the highest slope in (H, K, W1, W2) bands to the lowest in the W4 band. It's clear that the disk decay rate is slower in longer wavelengths, and the dust particles at larger radii evolve at a slower pace. The slope in the W3 band is also in accordance with those obtained from RBS2015 (Figure \ref{fig:log_plot} top right panel and equations \ref{eq:logw3} \& \ref{eq:logIM}). We did not include the short wavelength sources in RBS2015 as it consists of 3.4-4.6 $\mu$m, whereas our short wavelength encompasses a much broader range of 1.6-4.6 $\mu$m.

Panel 4 displays the result of our age-binned analysis, as explained
in Section \ref{sec:age bin}. Here, we have also fitted a linear relation to the disk fractions and age in the log scale. Similar to the previous analysis, the slopes are decreasing as the wavelength increases. The W3 and W4 slope values are comparable between both methods, demonstrating consistency in the longer wavelength evolution. However, we observe a difference in the H, K, W1, W2 slope (-1.94 in binned analysis of Section \ref{sec:age bin} vs. -1.17 in cluster-based analysis of Section \ref{sec:dis1} ), likely because short-wavelength disk fractions drop more rapidly after 10 Myr, and the binning process reduces the statistical influence of individual outliers that might moderate this decline in the cluster-based analysis. Despite this quantitative difference, both analyses confirm the overall wavelength-dependent pattern of disk evolution.

Our study expands the observational reach by adding a larger sample of clusters across a wider age range, complementing RBS2015. Our dataset has 18 clusters older than 20 Myr, whereas RBS2015 had 5 clusters. We also extend the observations to longer wavelengths (22-24 $\mu$m) for clusters up to $\approx$ 90 Myr, allowing us to study disk evolution in more evolved stellar populations. The discovery of disks in these older clusters provides fresh information on the physical processes influencing disk lifespan, which may include lower stellar mass (Section \ref{sec:massdisk}) and reduced photoevaporation \citep{2004ApJ...611..360A,2020MNRAS.496L.111C,2022ApJ...935..111L}. Longer wavelengths allow us to probe cooler outer disk areas where longer dynamical timeframes and lower incoming star radiation allow gas and dust to persist. The shallower slope seen at these wavelengths implies a slower rate of grain formation and settling, which might lead to a more gradual dispersal of the outer disk material.

\cite{2024AJ....168..159L} recently determined the disk fractions of stellar associations with ages up to 50 Myr, providing an important comparative dataset for our analysis. Our disk fraction values show strong agreement with Luhman's findings for W3 excesses across the 10–50 Myr age range and for W4 excesses in younger associations (10–30 Myr). However, a difference emerges in the 30–50 Myr range, where our average disk fraction for W4 exceeds that reported by Luhman. This discrepancy may stem from the relatively smaller number of individual cluster members in our sample compared to Luhman's more extensive dataset, as well as methodological differences in our respective analyses. Interestingly, Luhman discovered a non-monotonic evolution pattern for M0–M6 stars, with W3 excess fractions initially declining with age, reaching a minimum and then followed by a peak at $\approx$ 34 Myr, before declining again in older (40–50 Myr) associations. Our results reveal a remarkably similar non-monotonic trend, with disk fractions reaching a minimum value at 17 Myr, then increasing to higher values in the 20-30 Myr age range, before declining again in clusters older than 40 Myr. This consistent observations of a secondary peak in disk fraction across independent samples reinforces the possibility of a significant evolutionary process occurring in this age range and calls for further research.

Recent ALMA studies have considerably enhanced our understanding of protoplanetary disks in the solar neighborhood, establishing a critical baseline for comparing disk evolution across diverse environments and ages. Notable studies such as the ALMA Surveys of Lupus \citep{2016ApJ...828...46A} and Orion Molecular Cloud \citep{2020ApJ...894...74B,2020ApJ...890..130T} have uncovered specific properties of young disk populations. These studies generally focus on star-forming regions between 1-3 Myr, where disk detection rates are high. For example, the ALMA study of Lynds 1641 in Orion discovered 89 of 101 targeted disks in the dust continuum, resulting in a notable 88$\%$ detection rate \citep{2021ApJ...913..123G}. In addition to these mm observations, we also found high disk fractions in the continuum-dominated W3 (12 $\mu$m) and W4 (22 $\mu$m) bands. These mid-infrared observations explore the warm inner regions of disks and serve as a vital complement to ALMA's cold outer disk studies, providing a more comprehensive picture of disk structure and history at diverse spatial scales. Such findings give important information on the initial conditions and early growth of protoplanetary disks.  Our examination of evolved clusters shows a considerable decrease in disk proportions in later phases, indicating ongoing disk dissipation and possible planet formation. Our results on older clusters add to ALMA studies of nearby young areas, providing a more comprehensive picture of protoplanetary disks' life cycle from early star formation to progressive declination in older populations.

\subsection{Older Disk Candidates}
\label{sec:olddisk}
\begin{figure}
\includegraphics[width=1\linewidth]{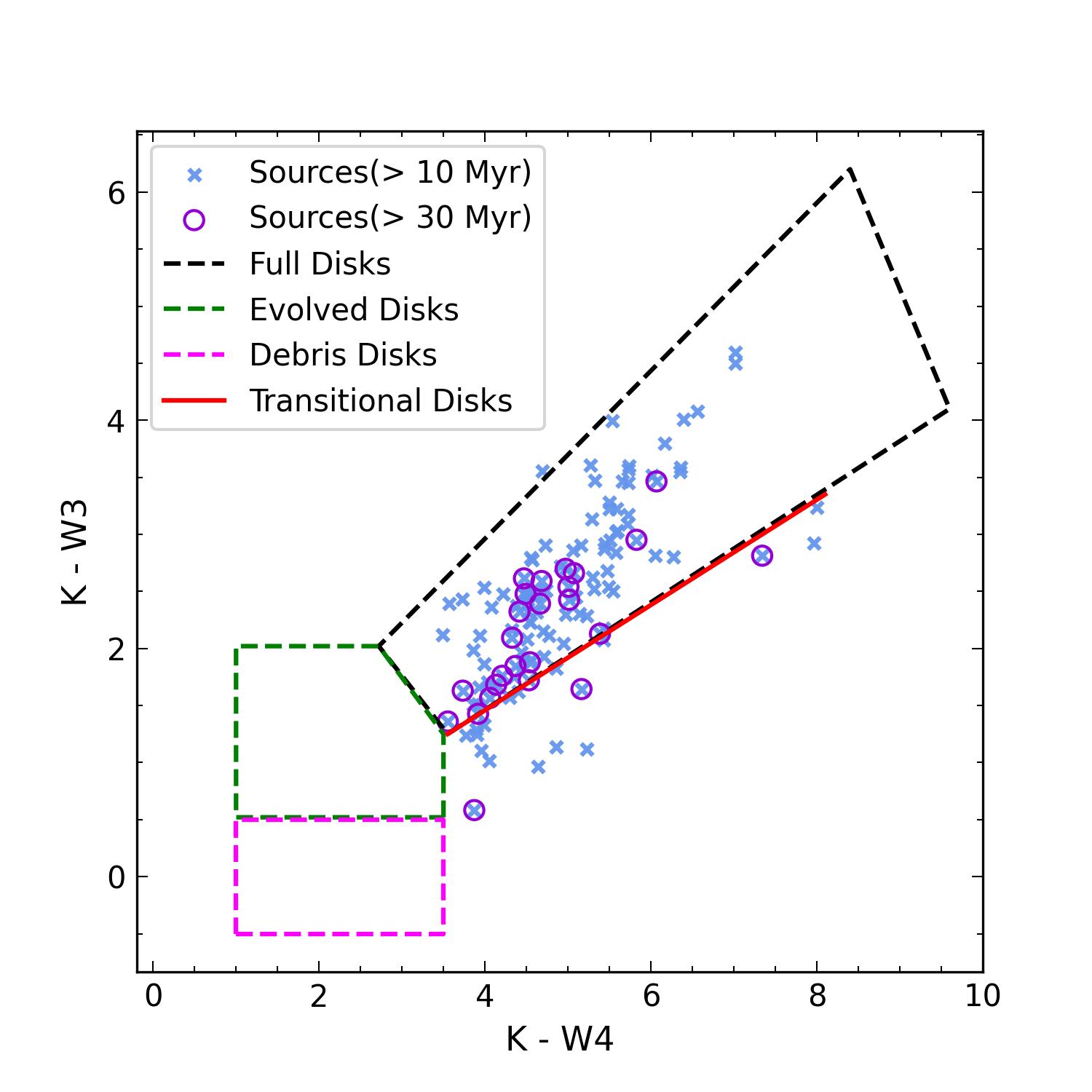}
\caption{K-W3 vs K-W4 CCD of the older disk candidates. Different regions indicates the different disk classifications from \protect\cite{2014ApJ...784..126E}
}\label{fig:w34ccd}
\end{figure}

Our sample encompasses not only young stars with full disks
but also stars beyond the typical time scale for disk dissipation
 (>10 Myr) that shows excess IR emission, suggesting the presence of long-lived disks. Using the K-W3 vs K-W4 CCD (Section \ref{dclass}), we can effectively classify these disk types. Four sources were removed from the initial sample of 124 older disk candidates due to their W3 and W4 band photometry being upper limits. These upper limits indicate that the sources are faint in these bands, and their flux measurements are likely affected by noise. The distribution of our older disk population is shown by the CCD in Figure \ref{fig:w34ccd}, where various regions denote different disk classes, according to \cite{2014ApJ...784..126E}. The figure shows that most of these sources are located in the full disk region, with lesser numbers displaying transitional disks. Among our 120 disk candidates, 95 sources fall within the age range of 10–30 Myr and the remaining 25 sources are older than 30 Myr  (indicated by violet circles).

The preservation of full disks at older ages suggests that primordial disk lifetimes were longer, a phenomenon  that might be connected to stellar mass effects, as discussed in Section \ref{sec:massdisk}. Although we attempted to study the accretion characteristics of these sources using existing LAMOST data (described in 
 Section \ref{sec:accr}), most of the sources lack corresponding LAMOST observations. Future spectroscopic observations will be required to determine their accretion characteristics. These older disk candidates provide valuable insight into the late stages of disk evolution and the diversity of dissipation timescales among protoplanetary systems.
 
\subsection{Mass of the disk sources}
\label{sec:massdisk}
The stellar mass exerts a considerable impact on the persistence and lifetime of circumstellar disks, with high-mass stars generally exhibiting shorter disk lifetimes compared to low-mass stars \citep{2006ApJ...651L..49C,2011ApJ...733..113R,2014MNRAS.442.2543Y,2015A&A...576A..52R,2022MNRAS.514.2315C,2023JApA...44...77D,2024ApJ...963..122P}.The main causes of this discrepancy are that stars with more mass have stronger stellar winds, faster accretion rates, and powerful radiation \citep{2005ApJ...630L.185C,2006A&A...459..837G,2008PhST..130a4024H}.  Because of these features, the disk material surrounding high-mass stars can erode and disperse more effectively over time, resulting in shorter disk lifetimes \citep{2014prpl.conf..475A}.

Furthermore, high-mass stars are more prone to disk fragmentation, which reduces the disk's lifetime \citep{2011ApJ...731...74B,2014prpl.conf..643H}. Rapid fragmentation leads to the consumption of disk material, which can then form planets or other bodies, so depleting it quickly. Low-mass star disks are more stable because of lower radiation levels, slower accretion rates, and reduced fragmentation. This greater stability enables disks surrounding
low-mass stars to exist for longer periods of time, contributing to their longer lifespans. To better understand the link between star mass and disk longevity, we examined how the mass range of sources with disks differed over age bins.

\begin{figure}
\includegraphics[width=0.9\linewidth]{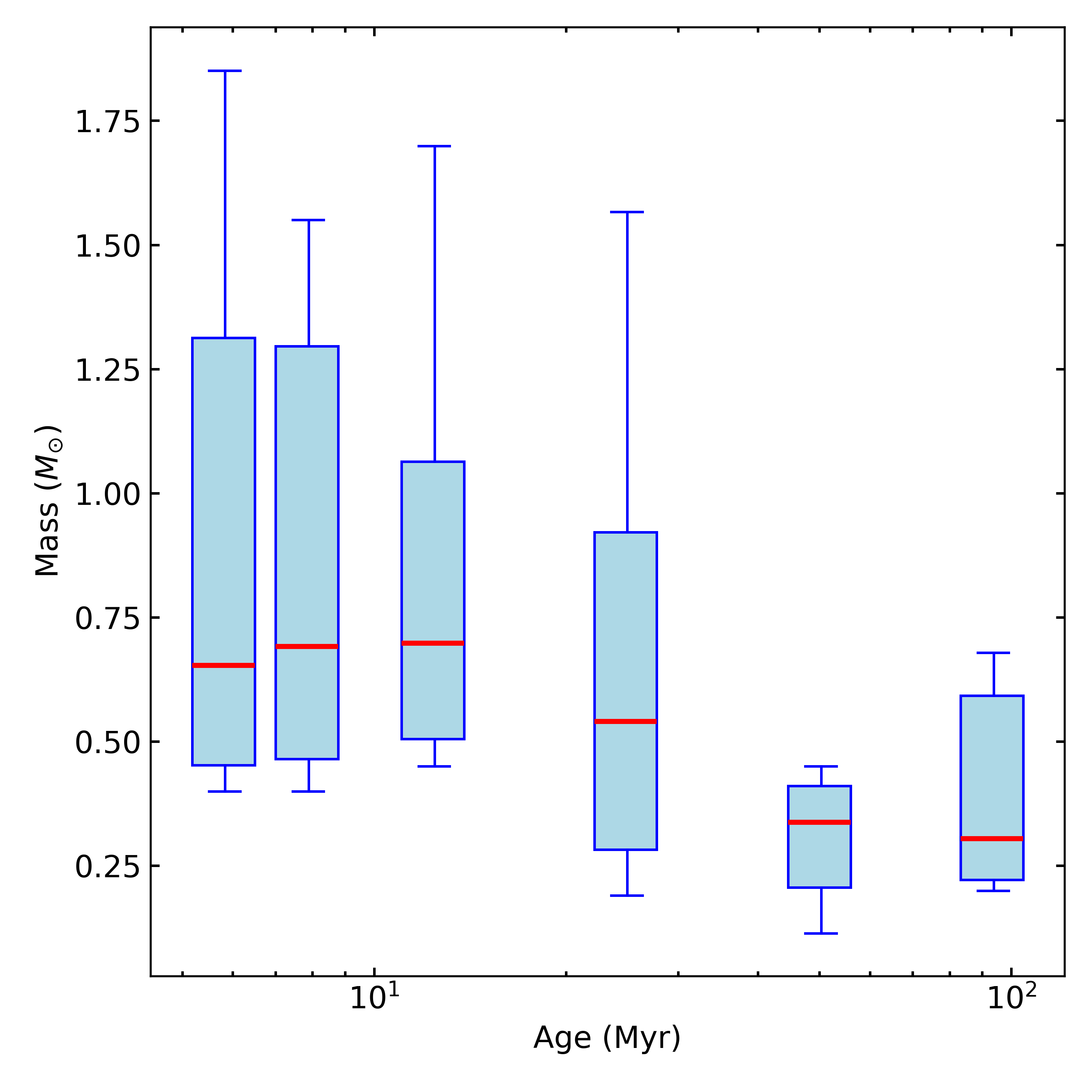}
\caption{Mass distribution of sources with disk around them in various age bins. The blue box indicates the 90\% mass range and the red line indicates the median mass for each bin. The whiskers indicates the entire mass range ($<$ 2 M$\odot$) of disk sources.
}\label{fig:medmass}
\end{figure}

The stellar mass distribution of disk-bearing sources as a function of the median age of the bins is shown in Figure \ref{fig:medmass}. The mass of these disk-bearing sources was determined by comparing their observed luminosity and temperature (derived from SED fitting) to PARSEC 1.2 isochrones, as explained in Section \ref{sec:param}. The disk-bearing sources considered here are from the total list, including excess sources from the analyses in H, K, W1, W2, W3 and W4 bands. We find a wide mass range of $\sim$ 0.4 - 2 \(M_\odot\) for disk-hosting stars in the younger populations ($\sim$ 1-10 Myr). Here the upper mass limit is due to the limit given in the source selection (see Section \ref{sec:massd}).  With increasing age bins ($>$ 40 Myr), this range substantially narrows down and shifts towards lower masses, with disk sources primarily falling within the mass range of $\sim$  0.1-0.75 \(M_\odot\). The red lines representing the median mass range for each age bin show a distinct drop in the median mass of the disk-bearing stars from 0.62 \(M_\odot\) in the youngest bin to 0.27 \(M_\odot\) in the oldest bin.

We acknowledge that our sample may be affected by selection effects, particularly for older clusters. Clusters with log(age) $>$ 7.6 ($\approx$ 40 Myr) are predominantly located within 200 pc, while younger clusters tend to be more distant (see Table \ref{table:main}). This distance-dependent sampling introduces a potential bias in our mass distribution analysis, which we try to minimize by assuming different stellar mass cuts for different distance ranges (Section \ref{sec:massd} ). The detection sensitivity for low-mass stars decreases with distance, potentially causing an under-representation of low-mass stars in distant (predominantly younger) clusters. Consequently, this could affect the precise determination of median mass values across age bins. Despite these selection effects, the consistent observation that only lower-mass stars ($<$0.75 \(M_\odot\)) retain disks at ages $>$ 40 Myr remains a robust finding. This finding is particularly significant as it aligns with theoretical expectations of mass-dependent disk dispersal mechanisms.

This pattern also suggests that lower-mass stars keep their disks for long periods of time. Lower-mass stars are expected to have longer disks due to decreased photoevaporation and stellar winds, according to theoretical models of disk development \citep{2009ApJ...696..143P,2021MNRAS.508.3611P}. Observational studies by \cite{2006ApJ...651L..49C} and \cite{2007AJ....133.2072D} revealed that while primordial circumstellar disks are retained by about 20 \% of K and M-type stars (0.1-1.2 \(M_\odot\)) in young clusters, higher-mass stars (above 1.2 \(M_\odot\)) show little to no evidence of such disks at infrared wavelengths, indicating a faster disk dispersal. Further supporting this trend, the disk fraction in Upper Sco is 5 \% for high mass stars (B7–K5.5-type) compared to 22\% for low-mass stars (M3.7-M6), and in UCL/LLC, the disk fraction is 0.7 \% for higher-mass stars compared to 9 \% for low-mass stars \citep{2022AJ....163...24L}.

The prior studies have focused on individual regions, making it difficult to understand the broad dependency of star mass on disk lifetimes owing to changing local environmental conditions \citep{2015A&A...576A..52R}. Our study tackles this issue by examining a wide range of stellar populations from 1 to 100 Myr across diverse environments. By combining a wide range of ages and doing an age-binned analysis, we overcome the limits of region-specific studies and present a more complete view of how disk evolution depends on stellar mass. The discovery of prolonged disk lifetimes for low-mass stars has significant implications for our knowledge of planet formation mechanisms and timescales. This shows that low mass stars may offer longer timeframes for planet formation. \citep{2021AJ....162...28V,2022ApJ...939L..10P}

\subsection{Accretors and Mass Accretion Rates}
\label{sec:macc}
\begin{figure}
\includegraphics[width=1\linewidth]{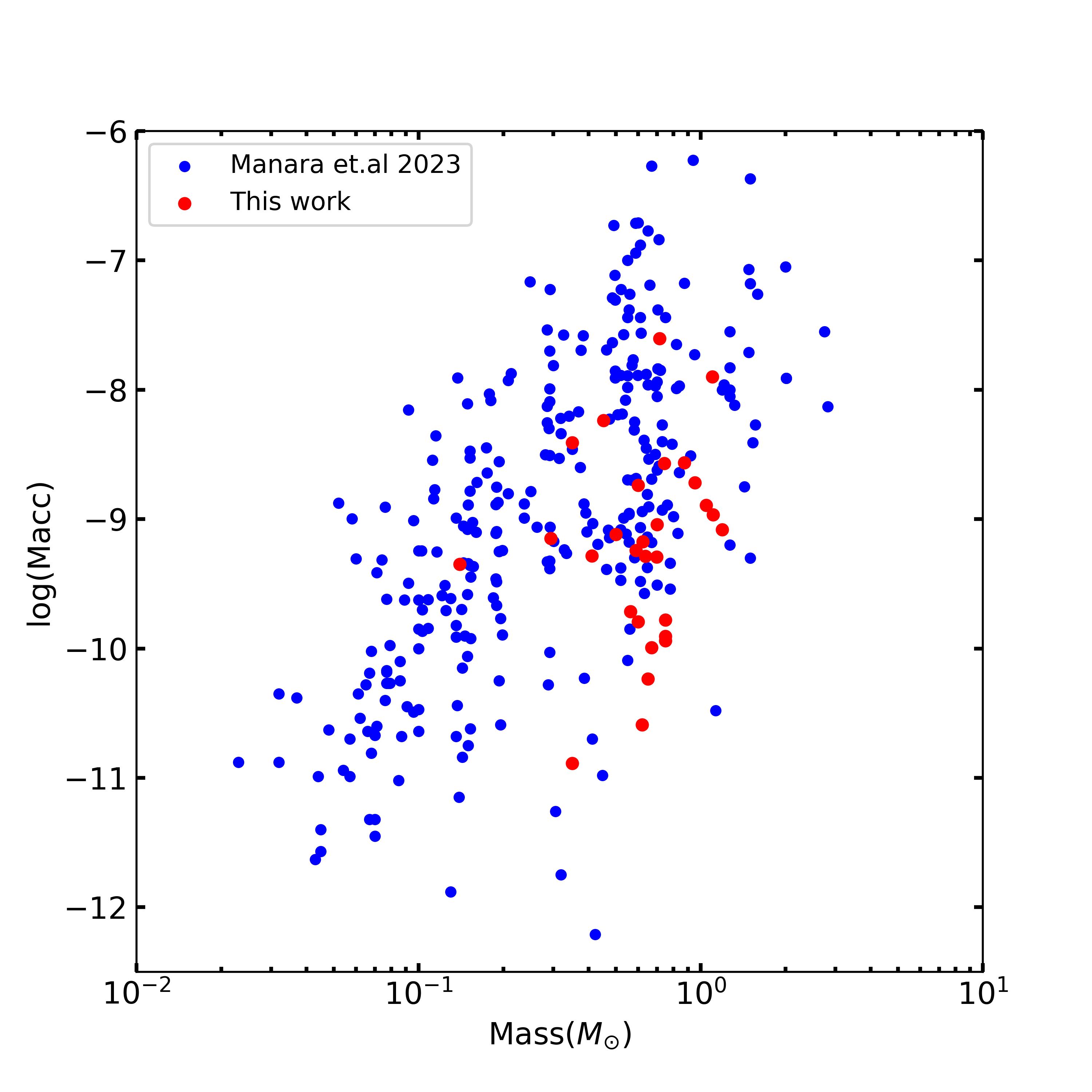}
\caption{Mass Accretion rate vs Stellar Mass plot. Red dots indicate the accretors in our sample. Blue dots are the accretors from Manara et al. 2023}\label{fig:acc_mass}
\end{figure}

The mass accretion rate of the 29 accretors was calculated as mentioned in Section \ref{sec:accr}. The mass accretion rates of the sources are in the range of $10^{-8} - 10^{-11} \, \mathrm{M}_\odot \, \text{yr}^{-1}$. \cite{2023ASPC..534..539M} have compiled a comprehensive sample of accretors from nearby young stellar clusters. Figure \ref{fig:acc_mass} shows the relationship between mass accretion rate and star mass for both our sources and those in Manara's sample. Our sample is mainly composed of more evolved clusters, where accretion rates are typically lower and less varied due to the depletion of disk over time, and our study includes a limited number of sources, which naturally limits the range of observed accretion rates and stellar masses. These factors account for our data set's narrower range. Furthermore, LAMOST counterparts were unavailable for the older disk candidates, which adds to the limitations of our data set’s coverage. 

We have spectroscopic data for clusters between 1-30 Myr and one cluster at 92 Myr, leaving a gap in our sampling between 30-90 Myr. Within this available data, we examined the ages of our accretors and discovered that all but one are less than 10 Myr. The 92 Myr old Melotte 22 cluster features one probable accretor, which is interesting, but high-resolution spectroscopy is necessary to confirm if it is a Peter Pan disk or other phenomena that mimics the accreting feature measured in LAMOST spectrum. This age distribution of accretors in our sample is consistent with theoretical models of disk development that predict a gradual reduction of disk material accessible for accretion onto the central star over time, leading to a decline in the accretion rate as the disk evolves \citep{2011ARA&A..49...67W,2014prpl.conf..475A}.  Our results also align with observational studies of accretion properties in nearby young clusters and star-forming regions \citep{2006A&A...452..245N,2012A&A...547A.104B,2014A&A...570A..82V,2017A&A...600A..20A,2021A&A...650A..43F}. These studies, spanning between 1 and 10 Myr, have consistently demonstrated a decrease in both the proportion of accreting stars and their accretion rates as stellar age increases.

\section{Conclusions}
We have analysed a sample of 6856 sources spanned across 32 young (1-100 Myr) clusters within 500pc. This study covers numerous previously unexplored clusters, which contributes to a better understanding of stellar and disk development in different environments. Our sample is based on Gaia-derived membership lists, which provide a comprehensive view of young stellar populations. SEDs were constructed using photometric data covering visible to mid-IR wavelength ranges, and the age and mass of the sources were determined using PARSEC 1.2 isochrones. The excess sources were identified using three criteria for the different IR wavelength ranges. Our aim was to understand how disk lifetimes vary in the evolved systems ($>$10 Myr). The main conclusions from our analysis can be summarized as follows:

\begin{enumerate}[label=\arabic*., align=left, leftmargin=*]
    \item Our analysis quantifies wavelength-dependent disk evolution, with characteristic timescales of $\tau_{\text{short}}$ = 1.6 ± 0.1 Myr for shorter wavelengths (1.6-4.6 $\mu$m) versus $\tau_{\text{W3}}$ = 4.4 ± 0.3 Myr for 12 $\mu$m, demonstrating a longer persistence in outer disk regions. This difference is further supported by our logarithmic analysis showing steeper slopes at shorter wavelengths (-1.17 for H, K, W1, W2) compared to longer wavelengths (-0.53 for W4). This indicates that dust particles at larger radii evolve more slowly since longer wavelengths correspond to larger radii.

    \item In contrast to shorter wavelengths where excess emission disappears by $\approx$ 20 Myr, we detect significant W3 (12 $\mu$m) and W4 (22 $\mu$m) band excess sources at older ages ($>$ 20 Myr) with disk fractions of $\approx$ 5-10 \% persisting to $\approx$ 50 Myr. We identified 120 disk candidates in clusters older than typical disk dissipation timescales ($>$ 10 Myr), with 25 sources older than 30 Myr, and the majority classified as full disks rather than transitional or evolved disks-suggesting that primordial disk structures can persist significantly longer than conventional models predict, though future spectroscopic observations are needed to verify their accretion properties and confirm their evolutionary status.
    \item Across all age bins, we found 33 transitional disk candidates. Transitional disks indicate the existence of an outer optically thick zone and an inner optically thin region, possibly implying ongoing planetary formation processes. This sample provides an excellent dataset for future spectroscopic studies, which are important for understanding this critical stage of planetary system evolution.
    \item Our analysis of disk-bearing sources across different age bins reveals a mass-dependent evolution pattern: the median mass of disk-hosting stars decreases from 0.62 \(M_\odot\) in the youngest clusters (1-10 Myr) to 0.27 \(M_\odot\) in the oldest clusters ($>$40 Myr), with the mass range narrowing from $\approx$ 0.4-2.0 \(M_\odot\) in young populations to $\approx$ 0.1-0.75 \(M_\odot\) in older populations. While selection effects related to distance-dependent sampling may influence the median mass values, the upper mass limit of 0.75 \(M_\odot\) for disk-bearing stars in populations older than 40 Myr remains a significant finding, providing strong quantitative evidence for theoretical models predicting longer disk lifetimes around lower-mass stars due to reduced photoevaporation and stellar winds.
    \item The mass accretion rates were computed using H$\alpha$ equivalent widths, demonstrating that most of the accretors in our sample are younger than 10 Myr. Our findings suggest that accretion rates decline with stellar age, which is consistent with earlier observations and theoretical models of disk development.
\end{enumerate}

\section*{Acknowledgements}
     
   This publication makes use of data products from the Two Micron All Sky Survey (a joint project of the University of Massachusetts and the Infrared Processing and Analysis Center/California Institute of Technology, funded by the National Aeronautics and Space Administration and the National Science Foundation), the Wide-field Infrared Survey Explorer (a joint project of the University of California, Los Angeles, and the Jet Propulsion Laboratory/California Institute of Technology, funded by the National Aeronautics and Space Administration) and the Guoshoujing Telescope (the Large Sky Area Multi-Object Fiber Spectroscopic Telescope LAMOST, which is a National Major Scientific Project built by the Chinese Academy of Sciences. Funding for the project has been provided by the National Development and Reform Commission. LAMOST is operated and managed by the National Astronomical Observatories, Chinese Academy of Sciences). This publication makes use of VOSA, developed under the Spanish Virtual Observatory (https://svo.cab.inta-csic.es) project funded by MCIN/AEI/10.13039/501100011033/ through grant PID2020-112949GB-I00.VOSA has been partially updated by using funding from the European Union's Horizon 2020 Research and Innovation Programme, under Grant Agreement nº 776403 (EXOPLANETS-A). GMB and
   JJ acknowledge the DST-SERB, Gov. of India, for the POWER  grant (No: SPG/2021/003850) for the financial support for carrying out this work. J.H. acknowledges support from the research projects, UNAM-DGAPA-PAPIIT IG-101723 and CONAHCyT grant 86372.

\section*{Data Availability}
The data underlying this article will be shared on reasonable request to the corresponding author.

\bibliographystyle{mnras}
\bibliography{references} 








\bsp	
\label{lastpage}
\end{document}